\RequirePackage{fix-cm}
\documentclass[smallextended]{svjour3}       % onecolumn (second format)
\smartqed  % flush right qed marks, e.g. at end of proof
\usepackage{graphicx}
\usepackage[T1]{fontenc}
\usepackage{listings}% http://ctan.org/pkg/listings
\usepackage{parcolumns}% http://ctan.org/pkg/parcolumns
\usepackage{seqsplit}
\usepackage{xcolor}
\usepackage{tikz}
\usepackage{multirow}
\usepackage{algorithm}
\usepackage{algorithmic}
\usepackage{url}
\usepackage{booktabs}
\usepackage{mdframed}
\usepackage{ragged2e}
\usepackage[utf8]{inputenc}
\usepackage{array}
\usepackage{amsmath}
\usepackage{afterpage}
\usepackage[multiple]{footmisc}
\usepackage{amssymb}

\usetikzlibrary{tikzmark,positioning}
\usepackage{blindtext}
\usepackage{pifont}% http://ctan.org/pkg/pifont

\usepackage{todonotes}
\definecolor{orange}{RGB}{255,102,0}
\definecolor{blue}{RGB}{0,153,255}

\usepackage{xcolor}
\newcounter{observation}
\renewcommand{\theobservation}{RQ\arabic{observation}}

\newcounter{subobservation}
\renewcommand{\thesubobservation}{\theobservation.\arabic{subobservation}}

\newenvironment{observation}
  {%
    \vspace{0.5\baselineskip}% Small space above
    \noindent% No indent
    \refstepcounter{subobservation}% Increase sub-observation counter
    \textbf{Observation \thesubobservation:}\ % Bold observation number and custom text
  }
  {%
    \vspace{0.5\baselineskip}% Small space below
  }

\def\BibTeX{{\rm B\kern-.05em{\sc i\kern-.025em b}\kern-.08em
    T\kern-.1667em\lower.7ex\hbox{E}\kern-.125emX}}

\begin{document}

% \title{ Smart Contract Upgradeability: An \\ Exploratory Study%\thanks{Grants or other notes
% %about the article that should go on the front page should be
% %placed here. General acknowledgments should be placed at the end of the article.}
% }
\title{Immutable in Principle, Upgradeable by Design: Exploratory Study of Smart Contract Upgradeability}

% \subtitle{Do you have a subtitle?\\ If so, write it here}


\author{Ilham Qasse         \and
        Mohammad Hamdaqa \and Björn Þór Jónsson%etc.
}


\institute{I. Qasse \at
              Reykjavik University, Reykjavík, Iceland \\
             \email{ilham20@ru.is}           %  \\
           \and
           M. Hamdaqa \at
              École Polytechnique de Montréal, Montréal, Canada
              \and
           B. Þ. Jónsson \at
              Reykjavik University, Reykjavík, Iceland
}

\date{Received: 01/12/2023}

\maketitle

\begin{abstract}
Smart contracts, known for their immutable nature to ensure trust via automated enforcement, have evolved to necessitate upgradeability due to unforeseen vulnerabilities and the need for feature enhancements post-deployment. This inherent contradiction between immutability and the necessity for modifications has prompted the development of upgradeable smart contracts. These contracts are immutable in principle yet upgradable by design, allowing for future updates without altering the underlying data or state, thus preserving the contract's original intent while accommodating necessary improvements.

This empirical study aims to bridge the gap in understanding the practical application and implications of upgradeable smart contracts on the Ethereum blockchain. By introducing a comprehensive dataset that catalogs the versions and evolutionary trajectories of smart contracts, the research explores several key dimensions: the prevalence and adoption patterns of upgrade mechanisms, the likelihood and actual occurrences of contract upgrades, the nature of modifications post-upgrade, and their impact on user engagement and contract activity. Through a detailed empirical analysis, this study systematically identifies upgradeable contracts and examines their upgrade history to uncover trends, preferences, and the practical challenges associated with applying such modifications.

The empirical evidence gathered from the analysis of over 44 million contracts shows that a mere 3\% embody upgradeable characteristics, with only 0.34\% of these undergoing subsequent upgrades. This finding underscores a cautious approach by developers towards contract modifications, possibly due to the complexity of upgrade processes or a preference for maintaining the original contract stability. Furthermore, the study demonstrates that upgrades are predominantly aimed at feature enhancement and vulnerability mitigation, particularly when the contracts' source codes are accessible. However, the relationship between contract upgrades and user activity is intricate, suggesting that additional factors significantly affect the utilization of smart contracts beyond their mere functional evolution.

\end{abstract}

\section{Introduction}

A smart contract is computer code deployed onto the blockchain that automates the enforcement, monitoring, and execution of agreements when predetermined conditions are met~\cite{wang2019blockchain,buterin2014next,sayeed2020smart,mohanta2018overview}. While initially envisioned to be immutable to ensure trust and reliability, acting as an unalterable agreement between parties, the practical application of smart contracts has revealed the necessity for upgradeability~\cite{he2020smart,qian2022smart,samreen2021survey,rodler2021evmpatch}. This realization stems from the dynamic nature of software, where vulnerabilities and bugs discovered post-deployment necessitate updates to ensure security and functionality~\cite{he2020smart,qian2022smart,samreen2021survey,sayeed2020smart,rodler2021evmpatch}. Thus, despite their foundational principle of immutability, the necessity for smart contracts to be upgradeable has become evident, highlighting the importance of modification capabilities to enhance functionality and address security concerns~\cite{antonino2022specification,salehi2022not,sayeed2020smart,rodler2021evmpatch}.

% \bj{$\downarrow$ This paragraph takes to long to make the point about maintaining the state.  This should be the first point.  After that, changes can be made to any part of the software, much like for traditional software.  (In traditional software, stored data can be updated, but I think this is typically a much more difficult operation than updating software, and hence is generally avoided.)}
% In contrast to traditional software, where a software upgrade can introduce changes to any part of the software, in smart contracts, a contract upgrade refers to the process of arbitrarily modifying the contract code logic while maintaining its stored data or state~\cite{antonino2022specification,salehi2022not}.
In contrast to traditional software, where a software upgrade can introduce changes to any part of the software, a smart contract upgrade refers to a process of arbitrarily modifying the contract code logic while maintaining its stored data or state~\cite{antonino2022specification,salehi2022not,bodell2023proxy}.
This requirement points to a key difference between smart contract upgrades and conventional software updates, where data preservation is less critical. An upgradeable smart contract is designed for possible future modifications by leveraging patterns that enable updates to its logic or features; hence, not all deployed smart contracts are upgradeable. Initially, the community relied on data migration to upgrade contracts, a method that, although effective in updating the contract's logic, did not align with the fundamental principle of preserving the contract's original state. This costly and complex approach led to the exploration of more efficient upgrading methods~\cite{antonino2022specification,salehi2022not,bodell2023proxy}.

In response to the limitations of the data migration approach, the community has introduced upgrading techniques, such as the proxy patterns~\footnote{\url{https://eips.ethereum.org/EIPS/eip-1822}}\footnote{\url{https://eips.ethereum.org/EIPS/eip-2535}}\footnote{\url{https://eips.ethereum.org/EIPS/eip-1967}}, which allows for modifying the contract's logic without affecting its state. These advancements have facilitated the creation of upgradeable smart contracts, yet there remains a gap in understanding how these approaches, especially the proxy pattern, are applied and their impact. The exploration of upgradeable smart contracts, particularly in assessing the prevalence of such contracts, how likely they are to be upgraded, the practical applications of upgrade patterns, and the impact of these upgrades on user engagement and trust, is still in its nascent stages. 

% \bj{$\downarrow$  The first sentence is a bit vague, esp. ``certain upgrading approaches'' -- you can be much more direct here.  }
% \bj{$\downarrow$  I think the decision to focus on Ethereum needs explanation.}
% This empirical study aims to identify upgradeable smart contracts that exhibit certain upgrading approaches and analyze the prevalence of these upgradeable contracts, including the commonly applied upgrading approaches. We focus on upgrading patterns introduced by the Ethereum community.\footnote{\url{https://ethereum.org/en/developers/docs/smart-contracts/upgrading/}} 
% % This empirical study aims to analyze real-world upgradeable smart contracts to provide evidence of their \MOE{What their refers to here, the upgradable contracts or the different types of upgrade patterns}{} prevalence and usage scenarios. 
% Moreover, we study the upgrade patterns through the lens of evolution lineage (different versions of deployed smart contracts) to answer the following Research Questions (RQs): 
This empirical study aims to provide an empirical perspective on smart contract upgradeability, addressing the research void with data-driven insights. It focuses on smart contracts deployed on the Ethereum blockchain due to its open-source nature and active community engagement in discussions and implementations related to smart contract upgradeability. The study collects existing smart contracts, focusing on their upgradeability features, to create a comprehensive dataset that catalogs smart contracts' versions and their evolution. Moreover, it analyzes the patterns of upgrades through the lens of evolution lineage (different versions of deployed smart contracts) to answer the following Research Questions(RQs):

\begin{itemize}
\item \textbf{RQ1: How prevalent are upgrade patterns in smart contracts?}

Although the smart contract community proposed many approaches to upgrading smart contracts, there is limited understanding of their adoption by developers and the prevalence of these methods in practice. Our research aims to fill this gap by designing Our research aims to fill this gap by designing policies that can be used as rules to identify upgradeable contracts and the approaches for their upgrades. These policies leverage regular expressions and code structure analysis, focusing on the proxy pattern family. We concentrate on seven community-endorsed standards, including the Universal Upgradeable Proxy Standard (UUPS)\footnote{\url{https://eips.ethereum.org/EIPS/eip-1822}} and Diamond (EIP-2535)\footnote{\url{https://eips.ethereum.org/EIPS/eip-2535}}, to analyze the adoption of upgradeable mechanisms systematically.

Our findings indicate that while upgradeable proxy contracts represent a small percentage of the total smart contracts analyzed, there is a clear preference among developers for certain upgrade pattern. This suggests a discernible trend in adopting specific standards, which has implications for the design and implementation of upgradeable smart contracts. The significance of this research lies in its contribution to understanding the current landscape of smart contract upgradeability.  Additionally, our research identifies a previously unrecognized types of upgradeable contract that diverges from established standards, suggesting innovative or customized approaches to upgradeability within the community. 
\item \textbf{RQ2: How likely is an upgradeable contract to be upgraded?}

Designing a contract to be upgradeable does not ensure it will be upgraded throughout its lifecycle. This question quantifies the extent to which contracts identified as upgradeable in our preceding analysis (RQ1) undergo actual upgrades. This exploration is critical for discerning the gap between the theoretical potential for contract evolution and the practical application of these upgrade mechanisms.
To address this, we trace the historical versions of smart contracts by analyzing logs and events associated with upgradeable contracts. This data is sourced from the Ethereum ETL dataset, focusing on upgrade requests and the subsequent new version addresses. We specifically look for emitted events that signal an upgrade and trace them to compile a comprehensive list of smart contract versions and their addresses. This approach allows us to systematically identify when upgrades have been implemented, shedding light on the actual frequency of upgrades among upgradeable contracts. 

Our findings reveal a notable restraint in the actual upgrading of contracts, with a minority of upgradeable contracts experiencing changes. This highlights a cautious approach to applying upgrades in practice, contrasting with the theoretical capability for frequent modifications. 
\item \textbf{RQ3: What changes occur in smart contracts post-upgrade?}

% \textbf{Motivation:} There is a lack of studies that analyze how upgrading approaches are applied in practice and the rationale behind upgrading a contract. 
% To answer this question, we analyze the evolution of smart contracts and identify the most common changes between smart contracts' versions. The results might help developers and researchers understand smart contracts' overall evolution and maintenance practices and make informed decisions about how to maintain and update contracts. 
% % \MOE{This may also help you uncover other patterns of upgrading}{}
Traditional software maintenance is well-established and classified into types such as corrective (fixing bugs), preventive (updating documentation, refactoring code), adaptive (updating for new environments), and perfective (adding new features, optimizing performance)~\cite{williams2010characterizing,chen2021maintenance,chapin2001types}. These classifications help in understanding the various motivations behind software updates and modifications. However, there is limited clarity on the application of these practices to smart contracts~\cite{chen2021maintenance}. In this research question, we aim to explore whether smart contract upgrades reflect these conventional maintenance activities, focusing on changes like bug fixes, feature modification, and efficiency improvements.

We analyze the smart contract versions identified in RQ2 to answer this question. This analysis involves a comprehensive review of code changes between versions, utilizing Git diff for code comparison and SmartBugs for security vulnerability detection. Additionally, we evaluate gas consumption alterations to identify instances of preventive maintenance through optimization efforts.

Through this process, we aim to categorize the post-upgrade changes into fixing vulnerabilities, feature modification (adding or deleting features), gas optimization, or residual ``other'' category for modifications that defy these conventional classifications. The results shows a significant portion of upgrades is categorized under ``other,'' indicating a variety of motivations beyond the primary ones identified.
\item \textbf{RQ4: How does smart contract upgrading impact the activity level of the contract?}

% \textbf{Motivation:} Smart contract upgradeability may lead to a trust concern for smart contract users. This is because many users trust contracts because of their immutability. However, smart contract upgradeability breaks the immutability of the contracts. Thus, this can reduce the number of users of the smart contract using it.
% Moreover, when a smart contract is upgraded, its internal logic and behavior may change, affecting how users interact with the contract. Users may need to adapt to changes in the contract's interface, and the gas consumption of transactions may also change. Additionally, if a smart contract is upgraded frequently or without proper communication to users, it may lead to a loss of trust and decreased activity.

% Analyzing the impact of upgrading a smart contract on its activity level enables developers and stakeholders to understand how upgrading a contract will affect its usage and overall adoption from its users. This can be useful for planning future developments and upgrades for the contract.
% Hence, in this research question, we analyze the impact of upgrading a smart contract on its activity level, considering each version's different lifetimes.
Smart contract upgradeability may lead to trust concerns among users, given the value placed on their immutability. The introduction of upgrades, altering a contract's internal logic or behavior, could necessitate user adaptation to interface or transaction gas consumption changes. Frequent upgrades or those not communicated effectively risk diminishing user trust and reducing contract activity.
Understanding the impact of upgrades on smart contract activity is crucial for developers and stakeholders to gauge how modifications influence usage and overall adoption. Therefore, we investigate the relationship between smart contract upgrades and their subsequent activity levels, taking into account the varied lifespans of each contract version.
We employ a detailed analysis of transaction volumes pre- and post-upgrade, extracted from Etherscan, adjusting for the contract versions' lifespan to ensure accurate assessment. This involves regression analysis to discern the effects of upgrades on contract engagement, providing insights into whether and how upgrades influence user interactions with smart contracts. 

The results demonstrate that most contracts (90.16\%) only recorded a single transaction, typically linked to their creation, indicating a prevalence of non-active contracts. Moreover, both linear and non-linear models exhibit limitations in capturing the relationship between upgrades and activity, suggesting that factors beyond lifespan and version number influence contract activity.
% \MOE{This is an okay question but looks foreign from the rest, you can keep it, I am just bringing your attention. Good studies are those that are sharp-focused}{}
\end{itemize}

% \bj{$\downarrow$ I think the first sentence should be before the RQs, and this paragraph could really be a summary of the contributions: ``In summary, this study makes the following contributions:'' followed by an actual list of items, which includes the dataset outcome (i) but then summarises the other outcomes in roughly items (ii) through (v) which recap the main outcomes of the four RQs.}
Our findings offer new insights into the nature and practical application of upgradeable contracts; a family of smart contracts that are \textit{immutable in principle yet upgradable by design}. In summary, the study's contributions are:
\begin{enumerate}
    \item We introduce an extensive dataset on smart contract versions, emphasizing their evolution. This dataset provides a valuable foundation for future research on smart contract upgradeability, offering insights into the development and maintenance of these contracts.
    \item Through an in-depth empirical analysis, we investigate the prevalence of upgrade patterns in smart contracts, identifying specific trends and preferences in adopting upgrade mechanisms. This analysis helps clarify the landscape of smart contract upgradeability, showcasing the diversity of approaches within the community.
    \item We quantify the likelihood of upgradeable contracts being actually upgraded, highlighting a cautious approach by developers towards modifying smart contracts. This contributes to understanding the gap between the potential for upgrades and their real-world application.
    \item Our study delves into the nature of changes implemented through smart contract upgrades, examining how these align with or diverge from traditional software maintenance practices. By categorizing these changes, we provide insights into the motivations behind contract upgrades.
    \item We analyze the impact of smart contract upgrades on user interaction and activity levels, discovering that upgrades do not necessarily lead to increased contract usage. This finding suggests that other factors, beyond the technical aspects of upgrades, significantly influence user engagement with smart contracts.
\end{enumerate}

% \bj{$\downarrow$ I like this paragraph, but it should be summarising things already said.}
% Key findings reveal that only 3\% of the 44 million smart contracts in our dataset are \textit{upgradable by design}. This could be due to the complexities involved or a preference for the reliability of non-upgradeable contracts. Moreover, only 0.34\% of these upgradeable contracts have been upgraded, indicating that real-world application of upgrades is far less common than the available capability. This disparity between upgradeable contract availability and their actual upgrades underscores a more conservative approach in smart contract development and maintenance, reflecting strategic decisions and potential complexities.

The remainder of this paper is organized as follows. Section~\ref{Bac} presents the principles of smart contracts and their upgradability. Section~\ref{SM} discusses the  research methodology utilised to answer the research questions. We present the results and discussion in Sections~\ref{Res} and~\ref{Dis}, respectively. Next, Section~\ref{RW} presents the related work while Section~\ref{TH} discusses threats to research validity. Finally, the paper is concluded in Section~\ref{Con}.

\section{Background}
\label{Bac}
This section introduces concepts related to this study's research questions, by providing an overview of smart contracts and their upgradeability.
\subsection{Smart Contract}
Smart contracts are computer programs that auto-execute agreements when certain conditions are satisfied. Smart contracts are mostly deployed on a blockchain network, enabling the execution of the contract to be secure and transparent ~\cite{wang2019blockchain,buterin2014next,sayeed2020smart,mohanta2018overview,wang2018overview}. 
Many blockchain platforms support smart contracts, such as Hyperledger Fabric~\cite{androulaki2018hyperledger}, Corda~\cite{brown2016corda}, and Ethereum~\cite{buterin2013ethereum}. This paper focuses on the smart contracts deployed on Ethereum due to the platform's open-source architecture and community. Ethereum is an open-source platform, meaning the underlying code and protocols are publicly available. Moreover, it has a large active community that supports researchers and developers with smart contracts. 

Ethereum runs its deployed smart contracts on the Ethereum Virtual Machine (EVM), a low-level stack machine that executes the compiled bytecode of the smart contract~\cite{buterin2013ethereum}.
Each operation in Ethereum requires a certain amount of computational effort, measured by gas. Gas is required for every operation in Ethereum, whether the operation is a transaction or the execution of a smart contract instruction. Some instructions are gas-intensive, such as instructions that utilize replicated storage. Gas metering prevents the sender from wasting computational power in executing unnecessary computation-intensive transactions. Moreover, it limits the number of instructions a transaction can execute, preventing non-terminating executions and DoS attacks.

Smart contracts in Ethereum are developed in many languages, such as Solidity, Vyper, and Bamboo. In this study, we target Solidity smart contracts, the most popular smart contract programming language~\cite{clack2016smart}. 
The Solidity language is an object-oriented language deployed on the EVM. In addition to Ethereum, several blockchain platforms support Solidity, including Quorum, Hyperledger Burrow, and Hyperledger Besu. Solidity has syntax similar to C and JavaScript, but it includes several unique concepts specific to smart contracts, including: visibility of function modifiers; internal, external, view; emitted events; and smart contract-specific operations such as self-destruct and revert.

We focus on Solidity contracts as there are more than 44 million Solidity contracts deployed on the Ethereum network. Additionally, Solidity supports libraries, which are useful for implementing reusable code in smart contracts. This might be beneficial in upgradeable smart contracts, as libraries can be modified and updated without requiring the redeployment of the entire contract.

\subsection{Smart Contract Upgradeability}
\label{SMU}
% \MOE{it would be nice here to have a figure such as the one in https://blog.chain.link/upgradable-smart-contracts/ also notice your definition is different, you don't need to follow their definition but the question is can you partially upgrade a contract?}{}
In smart contracts, upgradeability refers to the process of modifying smart contract code after deployment while maintaining the contract data and state ~\cite{antonino2022specification,salehi2022not}. 
Upgrading smart contracts has two benefits: (i)~it provides a mechanism to improve contract security by fixing security issues and bugs discovered post-deployment, and (ii)~it enables developers to add new features and functionality over time~\cite{antonino2022specification,salehi2022not}.

In the context of smart contracts, upgradeability and mutability (ability to change) are different. Since smart contracts are immutable by design, their code cannot be changed once deployed. However, the smart contract community proposed several mechanisms to upgrade the contract, such as deploying a new smart contract and directing user requests to the new one instead of the previous one. 

\begin{figure}[t!]
    \centering
    \includegraphics[width=0.8\columnwidth]{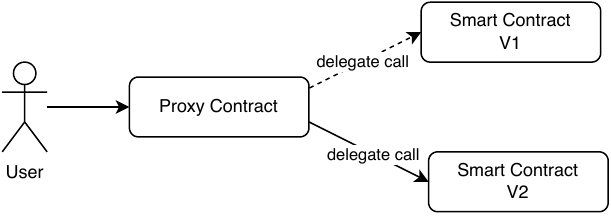}
    \caption{Proxy pattern for upgrading smart contract}
    \label{fig:proxy}
\end{figure}

The proxy pattern is a well-known method where users interact with a proxy contract, not directly with the business logic contract. The proxy contract holds data and forwards user requests to the designated smart contract version. When an upgrade is necessary, a new contract is deployed, and its address is updated in the proxy contract as the target version, as shown in Figure \ref{fig:proxy}. This approach allows for the seamless transition between contract versions without disrupting the user experience. Variants of this pattern include the Diamond Proxy and Universal Upgradeable Proxy Standard (UUPS), which maintain the core proxy concept but differ in their storage and code structure handling. The Diamond Proxy, for instance, allows for multiple logic contracts (facets) to be used simultaneously, providing greater flexibility and modularity.
The data separation pattern is another upgrading approach. It involves two contracts: one for storage and the other for business logic. Unlike the proxy pattern, users interact directly with the business logic contract, which interacts with the storage contract. This separation ensures that data remains intact and secure, even as the logic contract is upgraded or modified. The storage contract is a consistent data layer accessible only by authorized contract versions, ensuring data integrity and security.

Other upgrading alternative approaches include the strategy pattern and the data migration approach. The strategy pattern allows for the dynamic changing of algorithms or processes within a contract by separating the contract into interchangeable modules. The data migration approach involves transferring data from an older contract to a newer version. This process can be complex but allows for significant contract structure and logic changes.

In our study, we focus on the upgrading approaches introduced by the Ethereum community. We apply inclusion and exclusion criteria to select suitable approaches for this empirical study, as discussed in Section \ref{incl}.
\section{Study Methodology}
\label{SM}
\begin{figure}[hbtp]
    \centering
    \includegraphics[width=\textwidth]
    {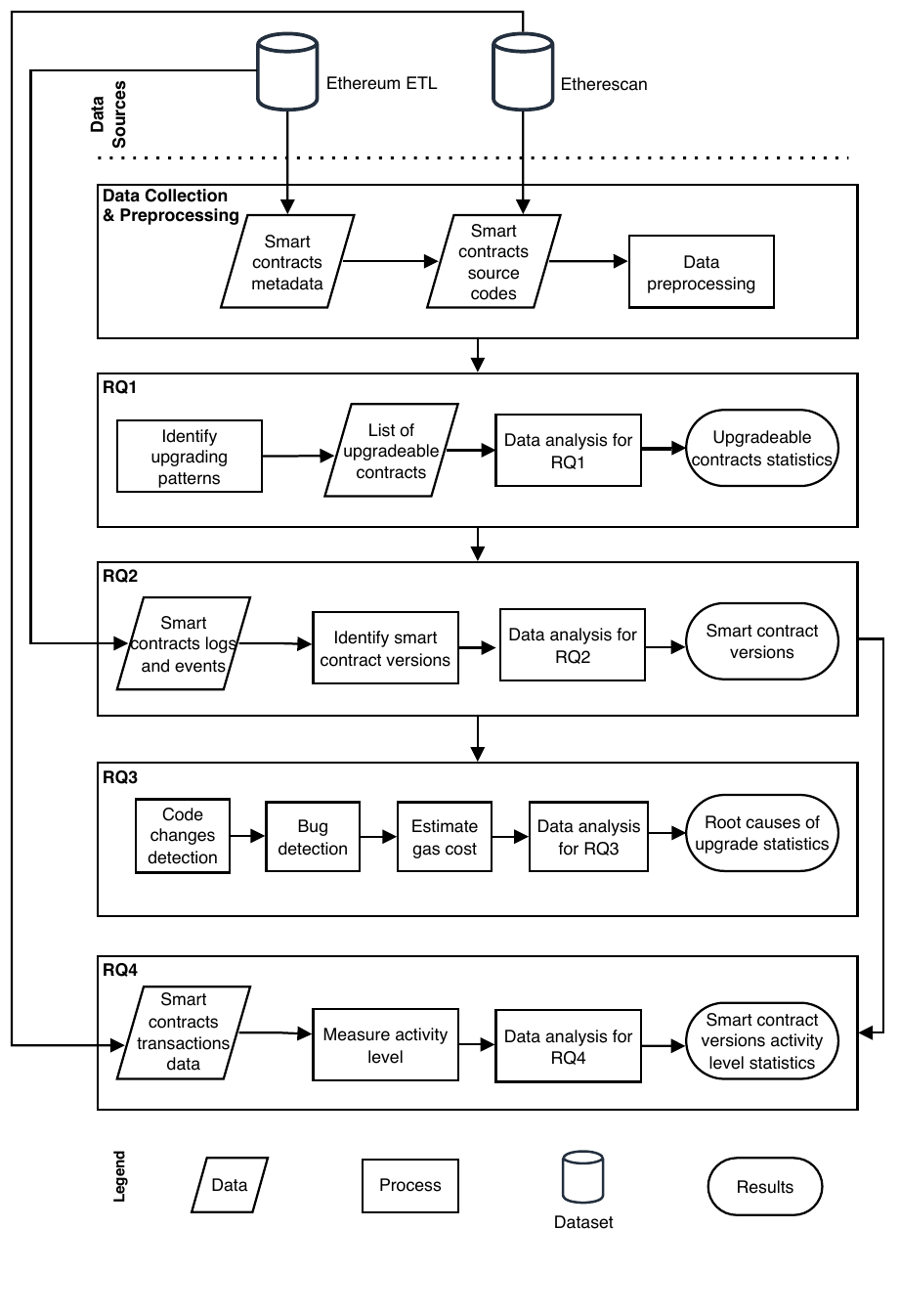}
    \caption{Overview of the study methodology}
    \label{fig:RO2}
\end{figure}

This study aims to empirically mine the upgradeable smart contract patterns from the deployed smart contracts to provide evidence of their prevalence and usage scenarios. Figure \ref{fig:RO2} illustrates an overview of our study methodology, including data collection, data preprocessing,  identifying upgradeable smart contracts, and analyzing smart contracts version evolution and their impact on the contract's activity level.
\subsection{Data Sources}
\label{DC}

In our study, we
integrate data from two sources, which are Ethereum ETL,\footnote{\url{https://ethereum-etl.readthedocs.io/en/latest/}} and Etherscan.\footnote{\url{https://etherscan.io}}
% \MOE{The dataset is called Ethereum in BigQuery, you need to provide a bit of explanation of each dataset for example for this one you can fine details here https://cloud.google.com/blog/products/data-analytics/ethereum-bigquery-public-dataset-smart-contract-analytics}{}

Ethereum ETL is a public Ethereum data explorer that enables users to explore blockchain data such as blocks and transactions. All the data is published as a Google BigQuery dataset.\footnote{\url{https://bigquery.cloud.google.com/dataset/bigquery-public-data:crypto_ethereum}} The Ethereum ETL dataset includes details about blocks, transactions, and smart contracts. Ethereum ETL configures nodes to synchronize data with the Ethereum blockchain, where these data are updated regularly. %We will
We extract from the dataset smart contract metadata such as smart contract address, bytecode, and the contract creator address. 

To obtain more details about the smart contract, such as the contract's source code and the contract's activity level, we use Etherscan, an Ethereum explorer. 
The data on Etherscan is updated in real-time, as it syncs with nodes configured on the Ethereum network. Based on each contract's address extracted from Ethereum ETL, we query Etherscan for the contract address and extract the contract's source code and activity level, if available. There are unverified smart contracts in Etherscan, which means that the developer didn't provide the source code of the smart contract. In this study, we do not consider these smart contracts.

\subsection{Data Collection and Preprocessing}
To answer the research questions, we preprocess the collected smart contracts data as follows:
\begin{itemize}
\item Remove comments and white spaces: This helps in reducing the size of collected data and normalizing these data by removing any inconsistencies in structure and format.
 % \item Group smart contract duplicates: According to the studies in ~\cite{pinna2019massive,pierro2020organized,oliva2020exploratory}, there are many duplicates in the deployed smart contracts. In this study, we will identify exact duplicates, group them, and consider only one representative to facilitate the analysis process of the source codes. This step is done after removing white spaces and comments.
 \item Reformat multi-file smart contracts: Smart contracts can be coded using one or multiple files. As we are comparing different versions of the smart contracts, it is essential to have each smart contract in a single file. For this purpose, we merge files and remove imports within files for multi-file smart contracts.
\end{itemize}
%\MOE{it would be nice if you can provide some reasoning why you need to do this}{}
% To facilitate analyzing smart contracts source codes, we will use the following preprocesses:
% \begin{itemize}
%     \item Removing comments and white spaces:  Although we have millions of verified contracts, most of these are duplicates~\cite{pinna2019massive, pierro2020organized, oliva2020exploratory}. Removing comments and white spaces is necessary to detect duplicated smart contracts.
%     \item Reformatting multi-files smart contracts: Smart contracts can be coded using one or multiple files. Since we will compare different versions of the smart contracts, it is essential to have each smart contract in a single file. For this purpose, we merged files and removed imports within files for multi-file smart contracts.
% \end{itemize}

% \begin{tikzpicture}[remember picture,overlay]
% \draw[red,rounded corners]
%   ([shift={(-3pt,2ex)}]pic cs:startc) 
%     rectangle 
%   ([shift={(3pt,-0.65ex)}]pic cs:endc);
% \end{tikzpicture}
We collected 44M deployed smart contracts. % up to July 2023. 
For each smart contract, we collected the following metadata: the smart contract address, the contract creator address, the timestamp, the compiler version, and the solidity version. Moreover, we collected from Etherscan the contract source code, where available, and the number of received transactions.

\subsection{Identifying Upgradeable Smart Contracts}
\label{incl}

% \bj{$\downarrow$ Moe proposed the following two paragraphs, but including footnotes inside todo notes causes the latex compiler to panic and give errors, so I made this edit... }
To answer RQ1, in our study on the prevalence of upgradeable smart contracts, we specifically focus on analyzing patterns that are widely recognized within the Ethereum,\footnote{\url{https://ethereum.org/en/developers/docs/smart-contracts/upgrading/}} and OpenZeppelin\footnote{\url{https://blog.openzeppelin.com/the-state-of-smart-contract-upgrades}} communities. We focus our analysis on patterns that offer clear, identifiable code and storage structures, enabling straightforward identification and differentiation. We exclude approaches that do not align with our criteria for systematic analysis. This includes ad-hoc upgrading methods such as Contract Migration, which often depend on manual interventions and lack clear distinguishable patterns in contract code. Similarly,  we excluded the Data Separation approach due to its minimal prevalence and adoption by the community, representing only 0.0007\% of the collected data \cite{salehi2022not}. Furthermore, We also omitted generic patterns focused more on design principles, like the Strategy Pattern, for their lack of direct relevance to upgradeable smart contracts.\footnote{\url{https://blog.openzeppelin.com/the-state-of-smart-contract-upgrades#strategy-pattern}} These exclusions are based on their limited relevance to our research objective that aims to shed light on common practices of upgradeability in smart contracts.

Accordingly, this work focuses on approaches that exhibit unique code features that enable us to establish differentiation criteria from other approaches (patterns). Notably, we concentrate on the family of proxy patterns that includes the Universal Upgradeable Proxy Standard (UUPS) and Diamond (EIP-2535), which exemplify the current best practices in smart contract upgradeability. These approaches are identifiable by specific features in their code, allowing us to establish policies that differentiate them from other approaches.

To detect upgradeable proxy contracts, we identify a set of policies that distinguish it from the other techniques. These policies include a combination of regular expressions specific to the upgrade pattern and its code structure. For example, the presence of \textit{delegatecall} is indicative of a proxy pattern that can be detected using a simple regular expression. However, not all proxy contracts are upgradeable contracts; forward proxy contracts are used to direct requests to other contracts and not to upgrade a smart contract. To distinguish these two types, it is necessary to analyze the code structure. Upgradeable proxy contracts usually include methods to upgrade the contract address, as shown in Listing \ref{lst1}. In contrast, the forward proxy does not have this method, and the requests are delegated to a fixed contract address, as shown in Listing \ref{lst2}.
\begin{lstlisting}[caption=Upgradeable proxy contract example,
  label=lst1, float]
pragma solidity ^0.8.0;

contract Proxy {
    address public implementation;

    constructor() public {
        implementation = address(new Implementation());
    }

    |\tikzmark{startb}|function upgradeTo(address _implementation) public|\tikzmark{endb}| {
        require(msg.sender == msg.sender, "Only the owner can upgrade the contract");
        implementation = _implementation;
    }

    function execute(bytes memory _data) public {
        require(implementation != address(0), "Implementation contract not set");
        (bool success, bytes memory returnData) = address(implementation).|\tikzmark{starta}|delegatecall(_dat|\tikzmark{enda}|a);
        require(success, "Execution failed");
    }
}
\end{lstlisting}
\begin{tikzpicture}[remember picture,overlay]
\draw[red,rounded corners]
  ([shift={(-3pt,2ex)}]pic cs:starta) 
    rectangle 
  ([shift={(3pt,-0.65ex)}]pic cs:enda);
\draw[red,rounded corners]
  ([shift={(-3pt,2ex)}]pic cs:startb) 
    rectangle 
  ([shift={(3pt,-0.65ex)}]pic cs:endb);
\end{tikzpicture}

% \moe{Our analysis covers a range of key standards that the community has proposed and widely adopted, addressing various facets of proxy contract functionality and upgradeability. These standards define the technical requirements for contracts that can be upgraded and reflect the evolving consensus within the blockchain development community on how best to achieve contract flexibility, security, and longevity. By focusing on these endorsed patterns and standards, our study aims to provide a comprehensive overview of the state-of-the-art in upgradeable smart contract technologies, contributing valuable insights to both the academic field and practical application in blockchain development.}

%Furthermore, different upgradeable proxy types exist, such as Diamond proxy and UPPS. The code structure of these types, including the storage structure, differs, which enables us to identify them. 
Our analysis covers a range of key standards proposed by the Ethereum Improvement Proposals (EIP),\footnote{\url{https://eips.ethereum.org}} and OpenZeppelin\footnote{\url{https://www.openzeppelin.com}} communities. These standards detail the technical requirements necessary for upgradeable contracts and establish a unified terminology for our analysis. Moreover, this allows us to identify and describe the different upgradeability methods systematically. By focusing on these recognized patterns and standards, our study seeks to provide a detailed overview of current practices in upgradeable proxy contract approaches. Accordingly, this paper concentrates on the following approaches: 
\begin{lstlisting}[caption= Forward proxy contract example,
  label=lst2, float]
pragma solidity ^0.8.0;

contract Proxy {
    address public implementation;

    constructor(address _implementation) public {
        implementation = _implementation;
    }

    function execute(bytes memory _data) public {
        require(implementation != address(0), "Implementation contract not set");
        (bool success, bytes memory returnData) = address(implementation).|\tikzmark{startd}|delegatecall(_dat|\tikzmark{endd}|a);
        require(success, "Execution failed");
    }
}


\end{lstlisting}
\begin{tikzpicture}[remember picture,overlay]
  \draw[red,rounded corners]
  ([shift={(-3pt,2ex)}]pic cs:startd) 
    rectangle 
  ([shift={(3pt,-0.65ex)}]pic cs:endd);
\end{tikzpicture}
\begin{itemize}
    \item EIP-897: DelegateProxy~\cite{Araoz_Izquierdo_2018} is a standard interface to facilitate interactions with various proxy types. It explicitly differentiates between two proxy types: forwarder proxies, which primarily act as relays to other contracts, and upgradeability proxies, which are designed to enable the evolution of contract logic over time. 
    \item ERC-1538 : Transparent Contract Standard~\cite{Mudge_2018} introduces a transparent way to manage smart contract upgrades. It focuses on enabling the addition, removal, and modification of functions within a contract in a manner that is clear and traceable. This standard emphasizes transparency in contract upgrades, ensuring that changes are understandable and auditable, thereby enhancing the maintainability and adaptability of contracts over time.
    \item EIP-1967: Proxy Storage Slots~\cite{Palladino_Giordano_Croubois_2019} addresses the necessity of a standardized approach for tracking proxy contracts and their corresponding logic contract addresses. The lack of a common interface for accessing this information had previously been a barrier to developing tooling for proxies. EIP-1967 responds to this need by specifying distinct storage slots within a proxy's storage, aiming to streamline the identification and management of both the logic contract and the proxy's administrative control. EIP-1967 introduces two distinct types of storage slots for proxies: Proxy Logic Storage Slots and Proxy Beacon Storage Slots. The Proxy Logic Storage Slots are utilized to store the address of the logic contract that the proxy delegates call to. This standardization facilitates the easy location and updating of the logic contract in a proxy setup. On the other hand, Proxy Beacon Storage Slots are employed in Beacon Proxy patterns. They store the address of a beacon contract, which in turn points to the logic contract. This method allows for the simultaneous upgrading of multiple proxy contracts by updating a single beacon contract.
    \item EIP-1822: Universal Upgradeable Proxy Standard (UUPS)~\cite{Barros_Gallagher_2019} enhances the developer experience by incorporating upgradeability directly into the logic contract. Unlike other upgradeability patterns where the proxy contract triggers updates, the UUPS allows the logic contract to initiate its own upgrade. This inversion of control not only streamlines the upgrade process but also provides developers with the option to remove upgradeability features in future iterations. 
    \item EIP-2535: Diamonds or Multi-Facet Proxy~\cite{Nick_2020} presents a flexible, efficient approach for smart contract development. Unlike traditional single logic contract proxies, ERC-2535 allows a smart contract to use multiple logic contracts, known as facets, enabling a more modular and extensible structure. This design allows for adding, replacing, or removing functionalities (facets) without deploying a new contract. 
    \item The OpenZeppelin Proxy Pattern~\cite{OpenZeppelin_2018} reflects one of the initial efforts in addressing the challenges of proxy contract development, such as function selector clashes. OpenZeppelin, a pioneer in this domain since 2017, proposed this pattern to mitigate the issues that arise when a proxy and its logic contract have overlapping function signatures. The pattern facilitates call delegation to the logic contract based on the caller's status, distinguishing between administrative calls and regular user interactions. This distinction ensures that only non-admin calls are delegated, while admin calls are handled directly by the proxy or are rejected if the function does not exist.
\end{itemize}
 
\begin{figure}[hbtp]
    \centering
    \includegraphics[width=\textwidth]
{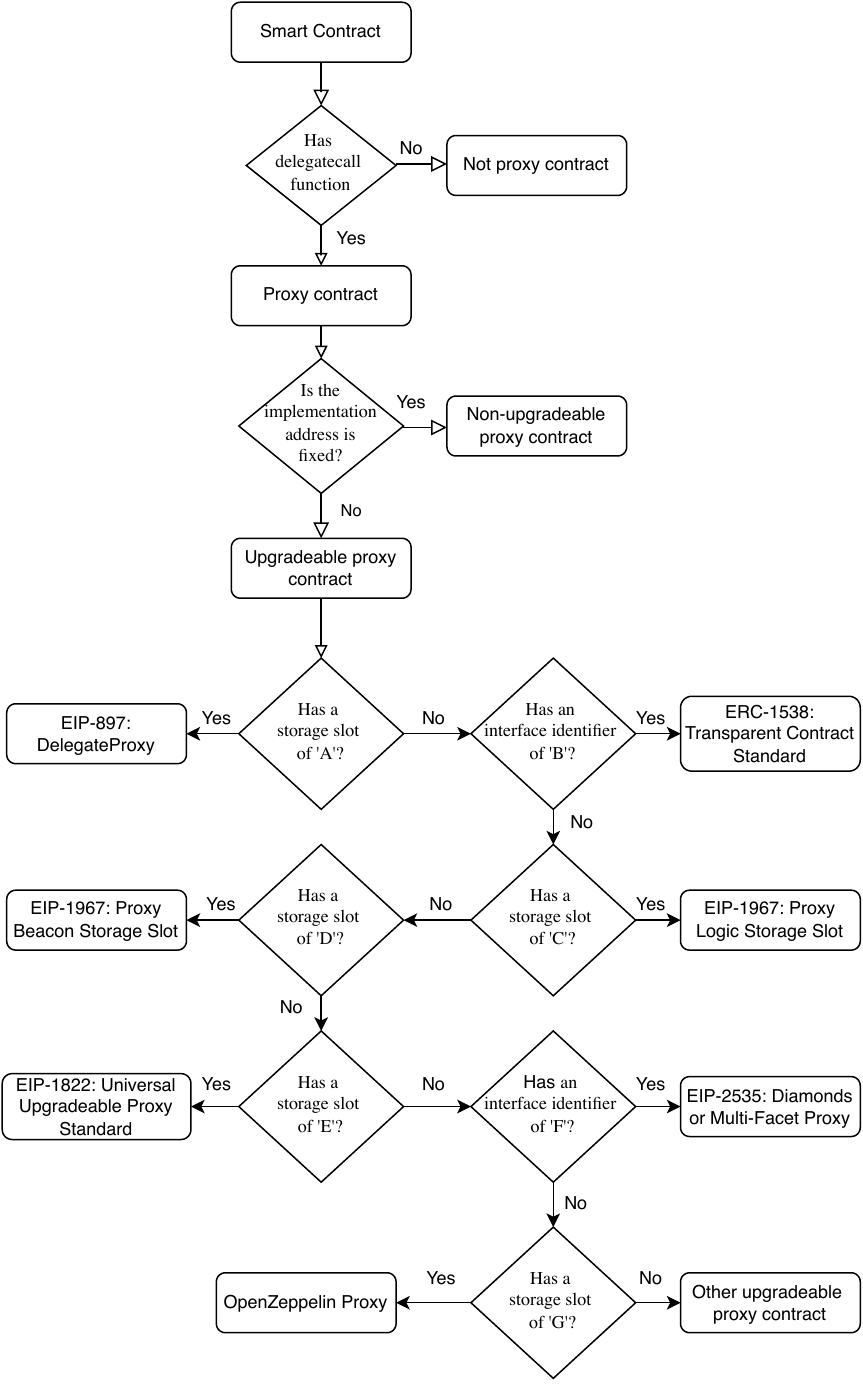}
    \caption{Flow chart to distinguish different  upgradeable proxy contracts}
    \label{fig:proxy1}
\end{figure}

We designed policies derived from the official standards specifications and utilized the Evm-Proxy-Identification tool,\footnote{\url{https://github.com/CaiJiJi/evm-proxy-detection/}} a well-known tool for detecting proxy contracts. However, due to the API constraints of the tool, we developed an independent proxy detection mechanism. This mechanism is based on the storage slots or unique identifiers (regular expressions) used in the standards, mirroring the approach followed by the Evm-Proxy-Identification tool. Figure \ref{fig:proxy} presents a flowchart of the process for identifying different upgradeable proxy contracts. If a proxy contract does not align with the identified types, we categorize it as \textit{Other Upgradeable Proxy Contract} to ensure comprehensive classification within our analysis, as shown in Figure~\ref{fig:proxy1}. Additionally, Table~\ref{tab:proxy}  lists the distinct storage slots or identifiers associated with each proxy type and their label in the flow chart.

\begin{table}[tb]
\caption{Proxy types and their unique identifiers}
\label{tab:proxy}
\resizebox{\textwidth}{!}{%
\begin{tabular}{@{}ccc@{}}
\toprule
\textbf{Label} & \textbf{Proxy Type}                                                                          &  \textbf{Storage Slot/Identifier}                                           \\ \midrule
A     & EIP-897: DelegateProxy                                                              &  ox5c60da1b00000000000000000000000000000000000000000000000000000000 \\ \midrule
B     &\begin{tabular}[c]{@{}c@{}}ERC-1538 : Transparent \\ Contract Standard\end{tabular} &  0x61455567 (Interface Identifier)                        \\ \midrule
C     & \begin{tabular}[c]{@{}c@{}}EIP-1967: Proxy Logic \\ Storage Slot\end{tabular}       &  0x360894a13ba1a3210667c828492db98dca3e2076cc3735a920a3ca505d382bbc \\ \midrule
 D     & \begin{tabular}[c]{@{}c@{}}EIP-1967: Proxy Beacon \\ Storage Slot\end{tabular}      & 0xa3f0ad74e5423aebfd80d3ef4346578335a9a72aeaee59ff6cb3582b35133d50 \\ \midrule
E & \begin{tabular}[c]{@{}c@{}}EIP-1822: Universal Upgradeable \\ Proxy Standard\end{tabular} &  0xc5f16f0fcc639fa48a6947836d9850f504798523bf8c9a3a87d5876cf622bcf7 \\ \midrule
F     &EIP-2535 : Diamonds Proxy                                                           &  interface ``IDiamondCut''                                            \\ \midrule
G     &OpenZeppelin Proxy                                                                  &  0x7050c9e0f4ca769c69bd3a8ef740bc37934f8e2c036e5a723fd8ee048ed3f8c3 \\ \bottomrule
\end{tabular}%
}
\end{table}
After identifying upgradeable proxy contracts, we statistically analyze how prevalent upgradeable proxy contracts are and what is the most commonly used upgrading pattern. By the end of this step, we have a list of upgradeable contracts, their address, and upgrading patterns.

\subsection{Historical Versions of Smart Contracts}
\label{SCHV}

\begin{figure}[t]
    \centering
    \includegraphics[width=\columnwidth]{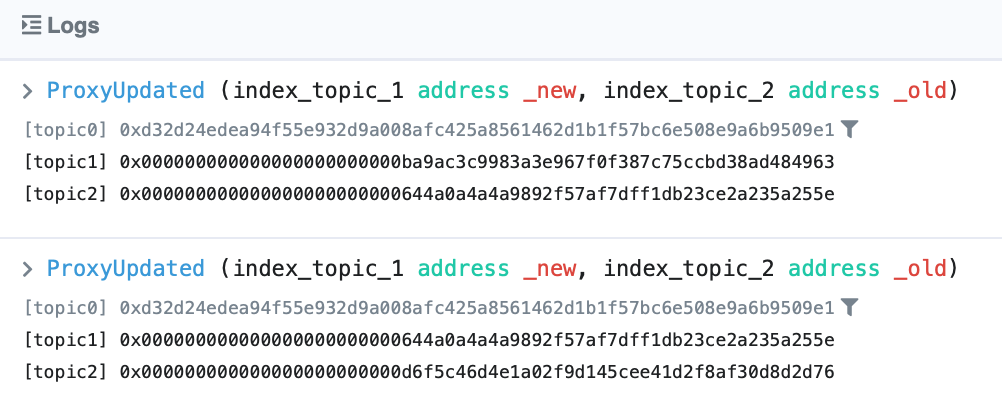}
    \caption{Sample of emitted upgrade contract events}
    \label{fig:event}
\end{figure}

There is no guarantee that all upgradeable contracts will be modified and upgraded. RQ2 aims to analyze how likely an upgradeable contract is to be upgraded.   Hence, we trace historical versions of each upgradeable smart contract, identified in RQ1,  to answer this research question. We trace versions by analyzing the logs and events of the upgradeable contracts. The logs and historical blocks for each upgradeable contract are obtained from the Ethereum ETL dataset. We identify the upgrade request (if available) from these logs and save the new version address and details.
For instance, some upgradeable contracts emit events with new contract addresses whenever there is a new upgrade for the smart contract. In these cases, we trace all the emitted events for the smart contract and extract the old and new version addresses.  Figure~\ref{fig:event} shows a sample of emitted events when a contract is upgraded obtained from Etherscan.

% To systematically identify and trace versions of smart contracts, this study focused on detecting upgrade requests and associated events within the contracts, utilizing logs to record detailed upgrade information. It's essential to acknowledge that upgrade mechanisms vary, influenced by the specific upgrade strategy employed. For example, certain proxy patterns, like the Diamond proxy, employ well-defined events as part of their standard for indicating upgrades. In contrast, other patterns might not use standardized event naming conventions, presenting challenges in event tracing due to dependency on precise event hashes, where a slight alteration in the upgrade event signature results in a completely different event hash.
% Our preliminary analysis examined upgrade events in a random sample of 5,000 proxy contracts, initially identifying 15 potential upgrade-related events. After a thorough evaluation, six events were deemed not relevant to our study objectives, culminating in a refined list of nine significant events:
Tracing the historical versions of smart contracts poses significant challenges, particularly in the context of event tracing. It is essential to acknowledge that upgrading event signatures vary and are influenced by the specific upgrade strategy employed. Moreover, not all contracts adhere to standard names for upgrading events. These variations in event naming conventions and signatures further complicate the tracing process. For example, a slight alteration in the upgrade event signature can lead to entirely different event hashes, rendering event tracing susceptible to discrepancies.

To address these challenges, we conducted a preliminary study to assess the existing event naming conventions, including those provided in standards such as the Diamond proxy pattern. We aimed to capture the event signatures commonly emitted when a smart contract is upgraded and understand the diversity of the signatures developers employ. We examined a random sample of 5,000 upgradeable proxy contracts and initially identified 15 potential upgrade-related events. Then we excluded six events from our analysis, as they were deemed irrelevant for tracing upgrade events and were primarily associated with unrelated activities. The remaining nine events considered for this study are:
\begin{itemize}
    \item Upgraded(address)
    \item NewImplementation(address, address)
    \item ProxyUpdated(address, address)
    \item FunctionUpdate(bytes4, address, address, string)
\item Upgraded(uint256, address)
    \item TargetUpdated(address)
    \item ImplementationUpdated(address)
    \item NewImplementation(bytes32, bytes32, address)
    \item ImplChanged(address, address)
\end{itemize}
These identified events facilitated tracing smart contract versions within proxy contracts and mapped them to the identified upgradeable proxies in RQ1. As a result, we have compiled a comprehensive dataset of upgradeable proxy contracts, their corresponding upgraded smart contract versions, and the upgrades they have undergone.
\subsection{Analyzing Post-Upgrade Changes in Smart Contracts}

% Identifying the Root Cause of Upgrade
% \MOE{the protocol in this step is not clear, It seems you are doing two different steps, one focusing on identifying code defs and the other focusing on studying if the original contracts had bugs? Do you think the title reflect this? Do you need more than one section/subsection for this? What is the goal here, is it Identifying the root cause of upgrade? }{}
The objectives of RQ3 are to identify and analyze the specific modifications undertaken in smart contract upgrades. This study explores how traditional software maintenance classifications, such as corrective and perfective, apply to smart contracts. The study seeks to ascertain and analyze the reflection of these established maintenance activities in the domain of smart contracts. In RQ3, we consider three post-upgrade changes of smart contracts: fixing vulnerabilities (corrective), feature modification (perfective), or optimizing gas cost (perfective). 
Identifying adaptive maintenance (updates necessitated by environmental shifts) and preventive maintenance (mitigating future issues) is not directly analyzed in this study due to inherent difficulties associated with its application and identification~\cite{chen2021maintenance}. Modifications that might otherwise fall under the adaptive or preventive categories or do not align with the aforementioned three post-upgrade changes are thus categorized under an ``other'' classification. 
In this step, we use the smart contract versions from RQ2 to analyze the root cause of their upgrade.

To identify the changes in post-upgrades, it is necessary to find the code changes in each smart contract version (upgrade). 
For code change detection between the identified smart contract versions (from RQ2),  we use Git diff,\footnote{\url{https://git-scm.com/docs/git-diff}}  a popular and powerful tool that is well-suited for comparing different code versions.

The Git-diff tool results help us locate the code changes for each upgrade. We analyze any security issues in the old and new versions to conclude whether the upgrade was to fix vulnerabilities. This analysis includes using SmartBugs~\cite{ferreira2020smartbugs} to detect vulnerabilities in smart contract versions. SmartBugs framework\footnote{\url{https://github.com/smartbugs/smartbugs}}  offers a wide range of security tools that cover different aspects of smart contract security, from static and dynamic analysis to formal verification and optimization. Using the combination of these tools enables us to analyze smart contract security.

Furthermore, we analyze whether there was gas optimization in the new versions of the contract. In this step, we estimate and compare the gas cost of the contract versions. The gas cost includes the gas fees for contract deployment and contract methods.

We follow these steps to identify the post-upgrade changes:
\begin{enumerate}
    \item Locate changes between smart contract versions. To identify the changes, we used the Git diff tool. 
    \item Run the smartBugs tool on both versions of the smart contract to detect any security vulnerabilities. If there is a security vulnerability in the first version that was not found in the second version, mark it as a bug fix.
    \item Identify any changes that were made between the two versions that do not address a security vulnerability. If the second version added lines of code that do not fix a bug, mark it as a new feature. Else if there were lines or functions removed without fixing any vulnerabilities in the code label it as a deleted feature.
    \item Compare the gas costs of both versions of the smart contract. If the gas cost has decreased in the second version, mark it as a gas optimization.
    \item For any modifications that do not fall into the categories of bug fix, new feature, deleted feature, or gas optimization, label these as ``other.''
\end{enumerate} Algorithm~\ref{alg:post_upgrade_changes} demonstrates the labeling process of post-upgrade changes based on the above steps.

\begin{algorithm}
\caption{Identifying Post-Upgrade Changes in Smart Contract}
\label{alg:post_upgrade_changes}
\begin{algorithmic}[1]
\REQUIRE $version1$, $version2$
\ENSURE $post\_upgrade\_changes$
\STATE $versions\_diff \leftarrow git\_diff(version1, version2)$
\STATE $vulnerabilities1 \leftarrow smartbugs(version1)$
\STATE $vulnerabilities2 \leftarrow smartbugs(version2)$
\STATE $bug\_fixes \leftarrow []$
\STATE $new\_features \leftarrow []$
\STATE $deleted\_features \leftarrow []$
\STATE $gas\_optimizations \leftarrow []$
\STATE $other \leftarrow []$
\FOR{$vulnerability$ in $vulnerabilities1$}
    \IF{$vulnerability \notin vulnerabilities2$}
        \STATE $bug\_fixes.append(vulnerability)$
    \ENDIF
\ENDFOR
\FOR{$line$ in $versions\_diff.added\_lines$}
    \IF{$line \notin vulnerabilities1$ \textbf{and} $line \notin vulnerabilities2$}
        \STATE $new\_features.append(line)$
    \ENDIF
\ENDFOR
\FOR{$line$ in $versions\_diff.removed\_lines$}
    \IF{$line \notin vulnerabilities1$ \textbf{and} $line \notin vulnerabilities2$}
        \STATE $deleted\_features.append(line)$
    \ENDIF
\ENDFOR
\STATE $gas\_cost1 \leftarrow calculate\_gas\_cost(version1)$
\STATE $gas\_cost2 \leftarrow calculate\_gas\_cost(version2)$
\IF{$gas\_cost2 < gas\_cost1$}
    \STATE $gas\_optimizations.append("Gas cost decreased")$
\ENDIF
\IF{len($bug\_fixes$) > 0}
    \STATE $post\_upgrade\_changes.append("Bug fix")$
\ENDIF
\IF{len($new\_features$) > 0}
    \STATE $post\_upgrade\_changes.append("New feature")$
\ENDIF
\IF{len($deleted\_features$) > 0}
    \STATE $post\_upgrade\_changes.append("Deleted feature")$
\ENDIF
\IF{len($gas\_optimizations$) > 0}
    \STATE $post\_upgrade\_changes.append("Gas optimization")$
\ENDIF
\IF{len($post\_upgrade\_changes$) == 0}
    \STATE $post\_upgrade\_changes.append("Other")$
\ENDIF
\RETURN $post\_upgrade\_changes$
\end{algorithmic}
\end{algorithm}

\subsection{Upgradeability Impact on Contract's Activity level} 
\label{SRQ4}
To analyze the impact of upgrading the contract on its usage (RQ4), we use the received transaction numbers for each version to get insight into the activity level. The transaction number is exported from the data collected from Etherscan. Since smart contract versions can have different lifespans, it is necessary to analyze the contract's activity level while considering its lifespan. The contract version lifespan is measured from when the contract was deployed until a new version was created. The age or lifespan is measured using the following equation:
\begin{align}
\label{eq:1}
  \mbox{\em lifespan}(V_a) = DT(V_{a+1}) - DT(V_a)
\end{align}
where  $V_a$ is the target version for which we want to calculate the lifespan $\mbox{\em lifespan}(V_a)$, $V_{a+1}$ is the next version, and $DT(V)$ is the deployment time of version $V$. If $V_a$ is the latest version, we consider the collection date instead of  $DT(V_{a+1})$ as the end of the deployment period.
To answer RQ4, we use a regression model with transaction numbers as the dependent variable and version lifespan as an independent variable. By including version lifespan in the model, we control for any differences in activity level due to the different lifespan of smart contract versions.  
Initially, we consider a linear regression model to examine if a linear relationship exists between the transaction count and the version lifespans. The equation can represent the linear regression model:
\begin{align}
\label{eq:2}
  \text{Transaction numbers} = \beta_0 + \beta_1 \times \text{Version lifespan} + \epsilon
\end{align}
where \(\beta_0\) is the intercept, \(\beta_1\)  is the coefficient for the independent variable representing the Version lifespan, and \(\epsilon\) represents the error term of the model.

% \moe{Did you consider complex models or not!!!}

% If a linear relationship is not sufficient to explain the data, we explore more complex models, such as random forest regression. Random forest regression is a machine learning technique that uses an ensemble of decision trees to make predictions. 
In our analysis, recognizing that a linear relationship might not adequately capture the complexities of our data, we have considered the application of random forest regression, a more complex machine learning model that uses an ensemble of decision trees to make predictions. We conduct an exploration of random forest regression to assess its effectiveness in our context. This technique allows us to capture non-linear relationships and interactions between variables that a simple linear model may overlook.
The random forest model does not have a simple equation like linear regression, as it is based on multiple decision trees that collectively contribute to the final prediction.
To assess the performance of our models (linear and random forest), we use the Mean Squared Error (MSE) and the R-squared (R²) score. The MSE is calculated as:
\begin{align}
\label{eq:3}
  \text{MSE} = \frac{1}{n} \sum_{i=1}^{n} (y_i - \hat{y}_i)^2
\end{align}
where \(y_i\)  is the observed value, \(\hat{y}_i\) is the predicted value by the model, and \(n\) is the total number of observations in the dataset.
The R-squared score, represented as:
\begin{align}
\label{eq:4}
  R^2 = 1 - \frac{\sum_{i=1}^{n} (y_i - \hat{y}_i)^2}{\sum_{i=1}^{n} (y_i - \bar{y})^2}
\end{align}
where \(\bar{y}\) is the mean of the observed values. The R² score measures the proportion of the variance in the dependent variable that is predictable from the independent variable(s). An R² score closer to 1 indicates that the model explains a large portion of the variance in the dependent variable.

Based on the calculated activity levels, we analyze the impact of upgrading on the activity levels of different versions of the contract.
\section{Results}
\label{Res}
This section presents the results of our empirical study, addressing the research questions and outlining our key findings. All used project source codes and the generated dataset are openly available in the project repository.
\subsection{RQ1: Prevalence of  Upgradeable Smart Contracts}
\afterpage{
\begin{figure}[t]
    \centering
    \includegraphics[width=\textwidth]{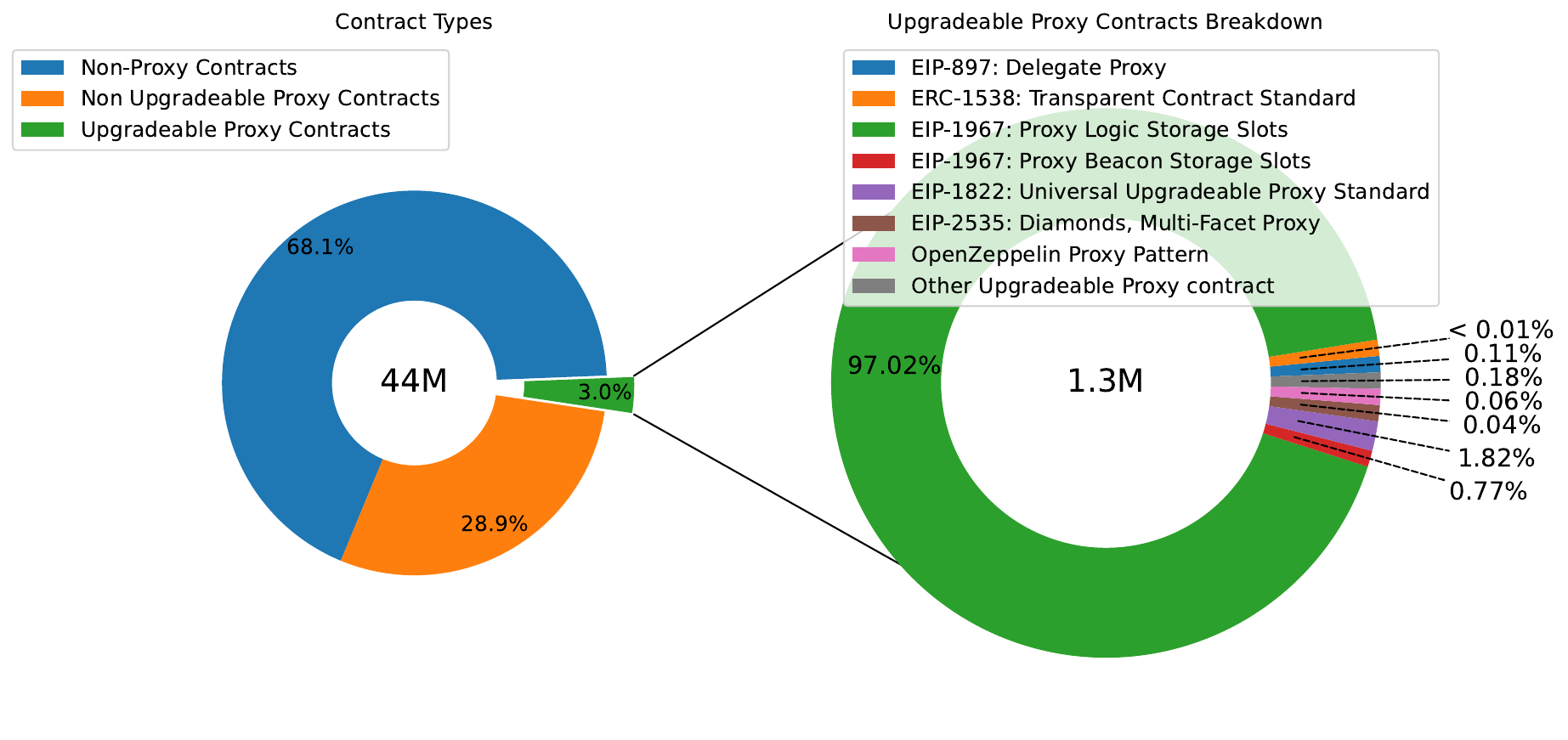}
    \caption{RQ1:Prevalence of upgradeable smart contracts}
    \label{fig:RQ1A}
\end{figure}}
Figure \ref{fig:RQ1A} presents a dual donut chart delineating the distribution of smart contract types in a dataset of 44 million contracts, focusing on upgradeable proxy contracts. The smaller donut chart categorizes these contracts into three distinct groups: Non-Proxy Contracts, Non-Upgradeable Proxy Contracts, and Upgradeable Proxy Contracts. The larger donut chart focuses on the Upgradeable Proxy Contracts segment, which provides insight into the distribution of different upgradeable standards mentioned in Section \ref{incl}.

\refstepcounter{observation}

\begin{observation}
\textbf{Upgradeable Proxy Contracts comprise only a small percentage, specifically 3\%, of the total dataset.} The largest segment, depicted in blue, represents non-proxy contracts, signifying their prevalence as the majority within the dataset. The orange segment, slightly smaller in size, corresponds to non-upgradeable proxy contracts. The smallest segment, depicted in green, is of key interest to our analysis as it represents upgradeable proxy contracts. 
The results from the chart indicate that Upgradeable Proxy Contracts account for only a small fraction of the total smart contracts quantified as 3\% of the dataset. This observation points to a limited adoption of upgradeable patterns in smart contracts. Despite their availability and potential advantages, the relatively small share of upgradeable contracts suggests that these patterns are not the primary choice in the smart contract domain. Factors influencing this include the complexity of implementing and managing upgradeable contracts or that developers are favoring the stability and predictability of non-upgradeable contracts over the potential benefits of upgradeability.
\end{observation}

\begin{observation}
\textbf{Among upgradeable contracts, \textit{EIP-1967: Proxy Logic Storage Slots} is the most utilized standard, suggesting a favored approach within the subset of upgradeable contracts.} Focusing on the Upgradeable Proxy Contracts segment, a secondary donut chart provides insight into the distribution of different upgradeable standards. Among these, the EIP-1967: Proxy Logic Storage Slots emerges as the most utilized, followed by other standards like the EIP-1822 Universal Upgradeable Proxy Standard and  EIP-1967 Proxy Beacon Storage Slots. This distribution within the upgradeable contracts highlights a preference for certain standards, underscoring the EIP-1967 Proxy Logic Storage Slots as the principal choice.
\end{observation}

\afterpage{
\begin{figure}
    \centering
    \includegraphics[width=\textwidth]{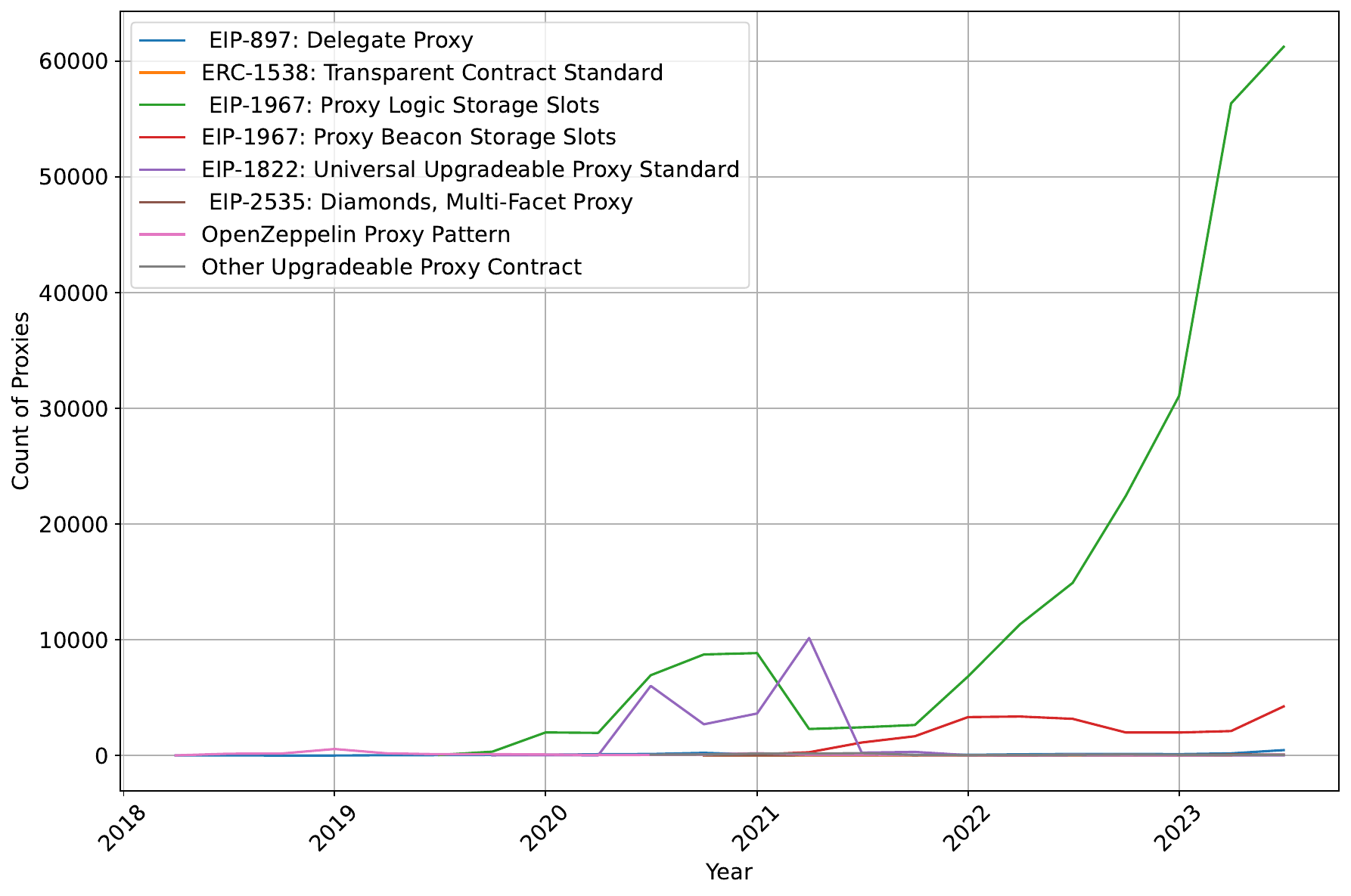}
    \caption{Time series analysis of the adoption of upgradeable proxy types}
    \label{fig:RQ1B}
\end{figure}}
\begin{observation}\textbf{Different proxies exhibit varying trends over time, highlighting dynamic shifts in technology adoption and standard preferences in the smart contract domain.} Figure \ref{fig:RQ1B} illustrates a nuanced landscape of proxy usage, delineating the counts of various proxy types over time up to mid-2023. In this figure, the x-axis represents the time dimension, which is segmented into quarters and provides a temporal context to the analysis. The y-axis quantifies the count of proxies, offering insight into the prevalence and adoption of different proxy types throughout the observed period. Each line corresponds to a different proxy type, their trajectories reflecting the evolving count of each across consecutive quarters. Notably, the chart encapsulates a spectrum of trends: while some proxy types exhibit a steady presence, others reveal dynamic patterns, including phases of decline, stability, or growth. An adjustment was made to account for partial data coverage in the last quarter represented in the data. Typically, each quarter comprises 90 days; however, the last quarter included only 15 days of data. To mitigate the potential distortion this discrepancy could introduce, the count for the last quarter was scaled to reflect a full quarter's worth of data. This scaling involved multiplying the observed count by a factor of 6 (i.e., 90/15), a methodological decision to preserve the integrity of the trend analysis.     

Delving into specific instances, the \textit{Proxy Logic Storage Slots (EIP-1967)} line presents an intriguing narrative. Initially, it shows a substantial presence, followed by a period of decline. This dip could indicate shifts in technology adoption, perhaps due to the emergence of more efficient or versatile proxy solutions. Concurrently, there is an uptick in the adoption of  \textit{UUPS (EIP-1822)}, suggesting a possible correlation. This period also coincides with the introduction of the \textit{Diamond Proxy (EIP-2535)}, which may have influenced the usage dynamics of other proxies.
However, it is worth noting that the decline in the usage of \textit{EIP-1967} was temporary, as there was a subsequent increase in its adoption. Interestingly, during this phase, \textit{EIP-1822} experienced a decrease in its usage, which could be due to the simplicity and efficiency of \textit{EIP-1967}, which might have regained favor over \textit{EIP-1822}. 
The \textit{Diamond Proxy (EIP-2535)} trend remains relatively stable throughout the observed period. This stability indicates a consistent, albeit niche, adoption. On the other hand, \textit{Delegate Proxy (EIP-897)}, one of the earliest introduced approaches, shows a trend of not being widely adopted over time. Its relatively low and stagnant usage suggests it may not have kept pace with evolving technological demands.
\end{observation}

\begin{figure}[t]
    \centering
    \includegraphics[width=\textwidth]{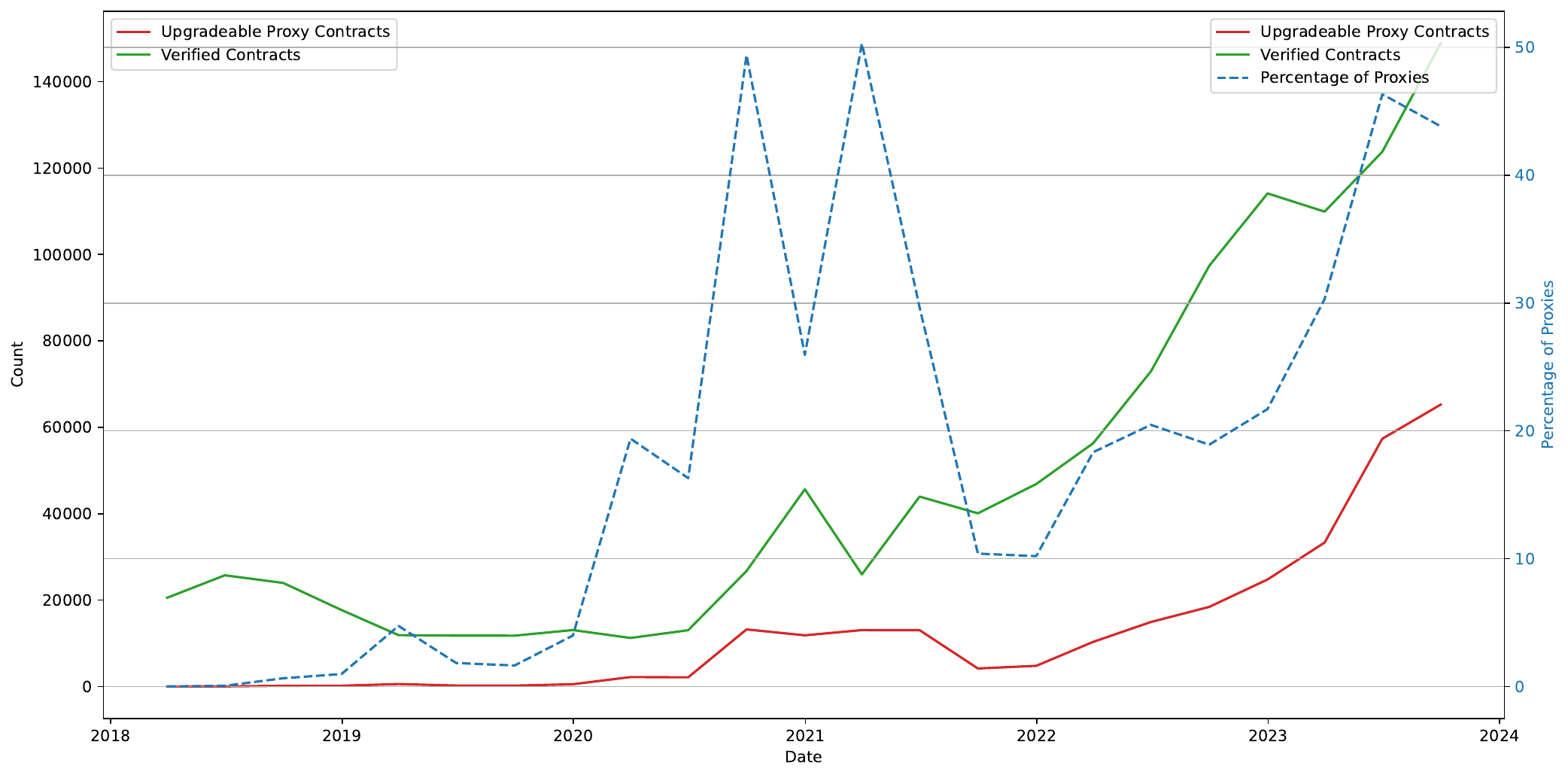}
    \caption{Trends in upgradeable proxy versus total verified smart contract deployments}
    \label{fig:RQ1C}
\end{figure}

\begin{observation}\textbf{There is steady growth in the deployment of upgradeable smart contracts over time.} Figure \ref{fig:RQ1C} presents the trends in upgradeable proxy contract deployments compared to the overall number of verified smart contracts over time. The x-axis outlines the timeline by quarters, illustrating the progression over several years. The primary y-axis on the left, colored in black, represents the count of contracts, tracking both upgradeable proxy contracts and total verified contracts. As in Figure \ref{fig:RQ1B}, the last quarter's data was scaled due to incomplete data.
A secondary y-axis on the right, highlighted in blue, details the percentage of upgradeable proxy contracts relative to the total verified contracts, offering insights into the adoption rate of upgradeable proxies within the broader smart contract ecosystem.
In Figure \ref{fig:RQ1C}, two distinct lines trace the evolution of contract counts: the red line for upgradeable proxy contracts and the green line for total verified contracts. These lines encapsulate the quantitative growth in each category, revealing patterns of increase, stability, or fluctuation over the observed period. Additionally, a blue dashed line marks the percentage of upgradeable proxy contracts, providing a relative measure of their prevalence against the backdrop of total contract activities.

In early 2018, when upgradeable proxy contracts were first introduced, there was no recorded use of them, highlighting their initial absence in the smart contract landscape. This period marked the beginning of an exploration into more flexible smart contract mechanisms beyond the traditional immutable contracts.

Following their introduction, the percentage of deployed upgradeable proxy contracts began to increase, indicating a growing interest and recognition of their potential benefits. This uptick in adoption showcased the development community's initial steps towards embracing the flexibility that upgradeable proxies offered for smart contract development.
The trend toward adopting upgradeable proxy contracts consistently rose from 2019 through the end of 2021. Despite experiencing fluctuations during this period, which likely reflected the community's ongoing experimentation and assessment of upgradeable proxies, the overall direction was clear: a steady move towards greater adoption. These fluctuations were part of the natural process of integrating new technology into existing practices as developers weighed the advantages of upgradeability against the foundational principles of blockchain technology.

Starting in 2022, the adoption of upgradeable proxy contracts witnessed a significant increase. This shift was marked by a broader recognition within the smart contract development community of the strategic importance of upgrading and improving smart contracts post-deployment. The surge in adoption during this period demonstrates a maturation in the understanding of upgradeable proxies, driven by the need for smart contracts to remain adaptable in the face of technological, regulatory, and operational changes. This phase underscores a pivotal movement towards prioritizing the adaptability and future-proofing of smart contract deployments.
\end{observation}

\begin{mdframed}[linewidth=1pt, linecolor=black]
\textbf{RQ1. How prevalent are upgrading patterns in smart contracts?} 

\vspace{.5ex}
\noindent\textbf{Answer:} The results show that Upgradeable Proxy Contracts constitute a mere 3\% of the dataset, suggesting a limited prevalence of upgradeable patterns in the smart contract landscape. This restrained adoption may stem from perceived complexities in implementing and managing such contracts or a preference for the stability of non-upgradeable contracts.
Despite this, the distribution of upgradeable proxy standards reveals a preference for EIP-1967, indicating nuanced selections within the upgradeable subset.
Further temporal analysis shows varying trends among proxy types, with some experiencing phases of decline and others gaining adoption, particularly noting the dynamic interplay between EIP-1967 and EIP-1822 standards. This reflects shifting preferences and technological advancements in the smart contract domain.
Moreover, a consistent increase in the deployment of upgradeable contracts from 2020 to 2021 signifies a growing developer inclination towards flexible and updateable contract structures. Despite a slight deceleration post-2021, the trend towards upgradeability remains upward.
Thus, while upgradeable proxy contracts initially represented a small fraction of the dataset, the evolving adoption patterns and preference shifts among developers suggest a gradual but discernible change in the landscape of smart contract development.
\end{mdframed}
\subsection{RQ2: Historical Versions of Smart Contracts}

To address RQ2, we focused on the number of smart contracts that were upgraded at least two times, based on the events discussed in Section \ref{SCHV}.

\refstepcounter{observation}

\begin{observation}
\textbf{Only a small fraction of upgradeable contracts have been upgraded.} From the dataset of 1.3 million upgradeable proxy contracts, a relatively small fraction, specifically 4,397, have been upgraded, resulting in 14,990 distinct smart contract versions. This observation underscores a general tendency within the smart contract community: despite the technical feasibility of frequent upgrades, the actual incidence of such upgrades is comparatively low.

\end{observation}

\begin{observation}\textbf{The \textit{Upgraded(address)} event is the most common, while others are rare, indicating different levels of standardization and application in upgrade processes.} Table \ref{table:upgrade_events} provides an overview of the prevalence upgrade events, with \textit{Upgraded(address)} being the most common, indicating its standard usage in the upgrade processes. In contrast, events such as \textit{ImplChanged(address,address)} are rare, signifying their specialized or less frequent application.
\begin{table}[t]
\centering
\caption{Frequency of different upgrade events in smart contracts}
\label{table:upgrade_events}
\begin{tabular}{c c}
\hline
\textbf{Event Type} & \textbf{Number of Upgradeable Proxies} \\
\hline
Upgraded(address) & 3,715 \\
NewImplementation(address,address) & 571 \\
ProxyUpdated(address,address) & 31 \\
FunctionUpdate(bytes4,address,address,string) & 21 \\
Upgraded(uint256,address) & 20 \\
TargetUpdated(address) & 19 \\
ImplementationUpdated(address) & 9 \\
NewImplementation(bytes32,bytes32,address) & 7 \\
ImplChanged(address,address) & 4 \\
\hline
\end{tabular}
\end{table}
\end{observation}

\begin{observation}
\textbf{A significant number of contracts undergo minimal upgrades, with the most common version number being 2.}
Figure \ref{fig:RQ2A} illustrates the distribution of versions across upgradeable smart contracts, revealing key insights into their upgrade patterns. The x-axis of the figure shows the number of versions, while the y-axis represents frequency. Each cross indicates the number of smart contracts with exactly that number of versions. 
The figure shows that the most frequent version number across these contracts is 2, highlighting a notable inclination towards at least a single upgrade cycle among these contracts. In contrast, the dataset reveals that the highest recorded version number is 70. Although such high version numbers are less common, as indicated by the decreasing density of crosses in the higher version range of Figure \ref{fig:RQ2A}, they point to some contracts being subject to extensive upgrading, undergoing numerous iterations over their lifecycle. Moreover, the average number of versions per contract is approximately 3.90, suggesting a general trend where most contracts are subjected to limited upgrades. This pattern is visually affirmed by the pronounced concentration of crosses at lower version numbers, suggesting that most contracts have fewer versions, indicating a trend toward minimal upgrades. As the version number increases, there is a noticeable decline in frequency, highlighting that higher version numbers are increasingly rare. 
\end{observation}
\afterpage{
\begin{figure}[t]
    \centering
    \includegraphics[width=\textwidth]
{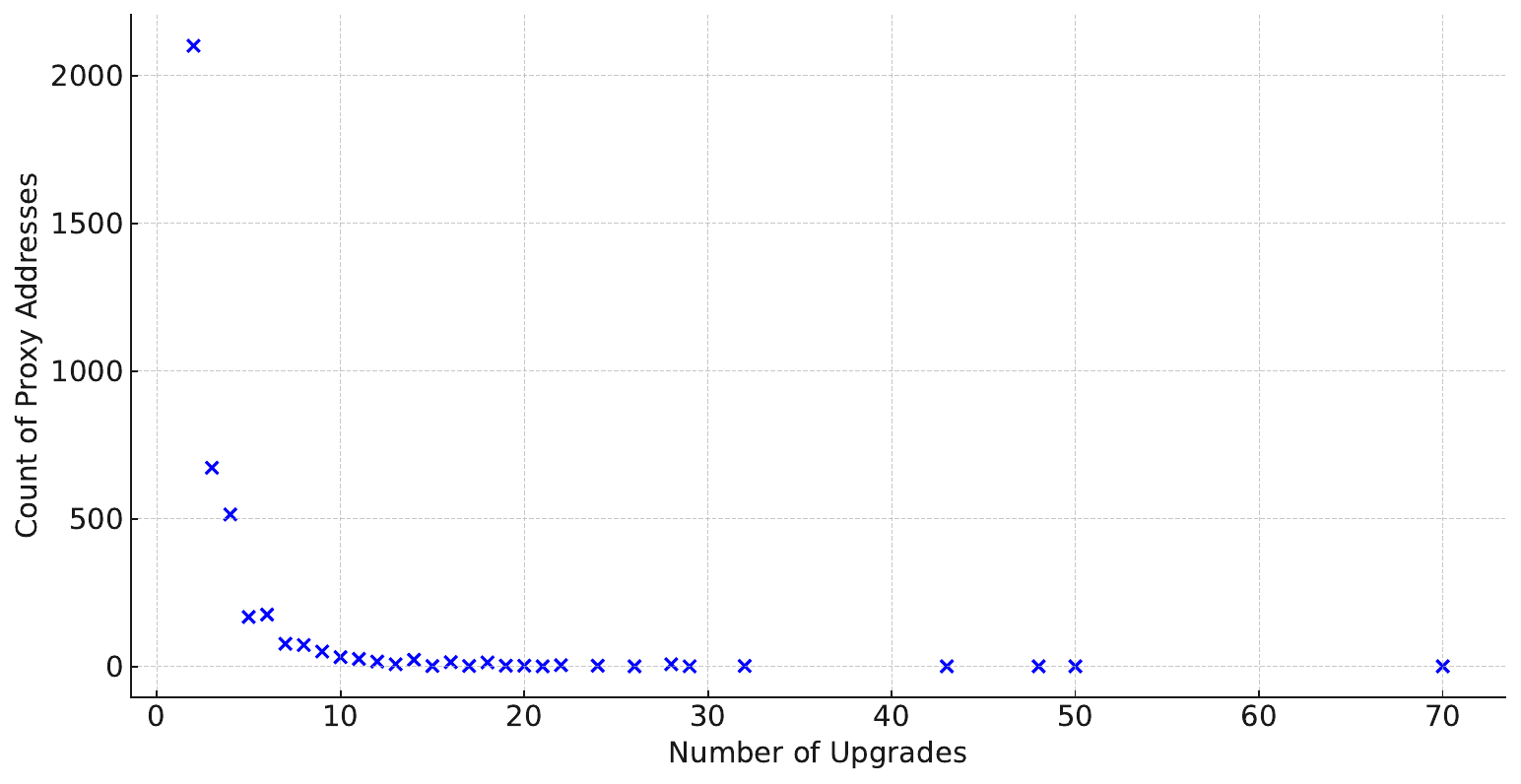}
    \caption{Historical smart contract versions (RQ2)}
    \label{fig:RQ2A}
\end{figure}}

\begin{observation}\textbf{The dataset reveals a notable replication trend in upgradeable smart contracts, with 50 unique versions being reused multiple times across various contracts.} Delving more into the details of the versions used in upgradeable smart contracts, we notice some interesting patterns. Specifically, when examining the dataset for duplicated versions, it is observed that 50 unique versions have been used more than once, either within the same contract (proxy address) or across different contracts. This indicates a practice of reusing specific versions, which could be due to various reasons, such as the popularity of certain contract versions or the reapplication of proven and stable implementations. For instance, the dataset analysis reveals that the most duplicated implementation address is \textit{0xf9e266af4bca5890e2781812cc6a6e89495a79f2}, which has been replicated 394 times across various contracts. 
\end{observation}

\begin{observation}\textbf{Analysis of the top 5 proxies indicates two primary upgrade strategies: a majority displaying unique versions for diverse upgrades and a minority favoring version duplication for stability and specific functionalities.} The existence of duplicates in smart contract versions draws attention to the need for a more focused analysis, particularly on proxies with a high number of versions. Table \ref{tab:top5} demonistrates the top 5 proxies, categorized by the highest number of versions, and examines the uniqueness of these versions.

The table shows interesting patterns about the upgrade strategies of these top five proxies:
Proxies \textit{\seqsplit{0x62faa8937f71b4896f9b250f675ff89a5f6875cc},  \seqsplit{0xb5c9985dc029b37d756938745760747b62ff46f9}}, and \textit{\seqsplit{0x2f6081e3552b1c86ce4479b80062a1dda8ef23e3}} show a high degree of uniqueness in their versions. The close correlation between the number of versions and the number of unique implementations suggests that most versions introduced are distinct, reflecting a strategy of consistent and diverse upgrades.
On the other hand, proxies \textit{\seqsplit{0x31946680978cefb010e5f5fa8b8134c058cba7dc}} and 
\textit{\seqsplit{0x1920d646574e097c2c487f69f40814f95d45bf8c}} display a different trend, with a significant portion of their versions being duplicates. This indicates a more conservative approach where certain versions are reused, possibly due to their stability or specific functionality that suits the contract's needs over time.
\begin{table}[t]
\centering
\caption{Top five proxies with number of versions and unique implementations}
\label{tab:top5}
\resizebox{\textwidth}{!}{%
\begin{tabular}{>{\centering\arraybackslash}p{0.6\columnwidth}>{\centering\arraybackslash}p{0.2\columnwidth}>{\centering\arraybackslash}p{0.2\columnwidth}}
\toprule
\textbf{\begin{tabular}[c]{@{}c@{}}ProxyAddress\end{tabular}} & \textbf{\begin{tabular}[c]{@{}c@{}}Number \\ of Versions\end{tabular}} & \textbf{\begin{tabular}[c]{@{}c@{}}Unique \\ Implementations\end{tabular}} \\
\midrule
0x2f6081e3552b1c86ce4479b80062a1dda8ef23e3 & 70 & 62 \\
0x62faa8937f71b4896f9b250f675ff89a5f6875cc & 50 & 25 \\
0x1920d646574e097c2c487f69f40814f95d45bf8c & 48 & 23 \\
0x31946680978cefb010e5f5fa8b8134c058cba7dc & 43 & 42 \\
0xb5c9985dc029b37d756938745760747b62ff46f9 & 32 & 31 \\
\bottomrule
\end{tabular}}
\end{table}

\end{observation}

\begin{figure}[t]
    \centering
    \includegraphics[width=\textwidth]
    {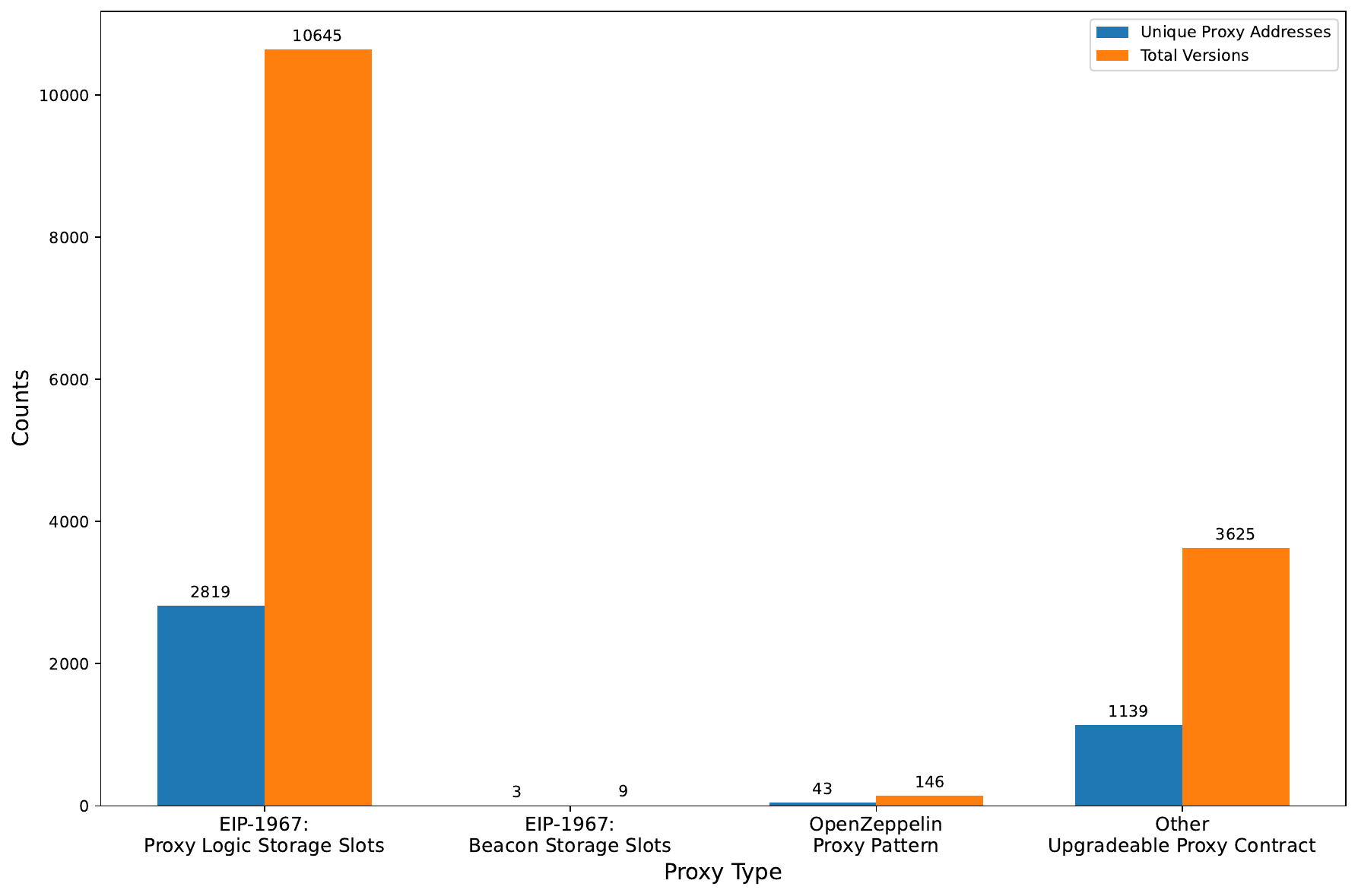}
    \caption{Smart contract versions versus proxy type}
    \label{fig:RQ2B}
\end{figure}

\begin{observation}\textbf{The study shows a dominant prevalence of the Proxy Logic Storage Slots (EIP-1967) in upgradeable smart contracts, with less standardized or custom hybrid proxy types also present but with lower adoption rates.} Regarding the analysis in the context of RQ1, where we study the prevalence of upgrading approaches, Figure~\ref{fig:RQ2B} illustrates the distribution of proxy types based on unique proxy addresses and their total versions. This visualization provides a clear comparative view of the proxy types' adoption and number of upgrades.
The Figure shows that existence of diverse range of less standardized or custom hybrid proxy implementations which is not following any identified standards. We classify these proxy types as a \textit{Other upgradeable proxy contract}. 
Similar to the results of RQ1, the \textit{Proxy Logic Storage Slots (EIP-1967)} type is the most predominant, both in terms of the number of proxy addresses and total versions. The \textit{OpenZeppelin Proxy Pattern} and \textit{Beacon Storage Slots (EIP:1967)} is reletively low. The missing patterns in Figure~\ref{fig:RQ2B} do not have any upgrades, indicating that these proxies were not used at all. An interesting observation is the missing upgrades using the approach \textit{UUPS (EIP-1822)} which was the second most used proxy standard in RQ1.
\end{observation}

\begin{figure}[t]
    \centering
    \includegraphics[width=\textwidth]
    {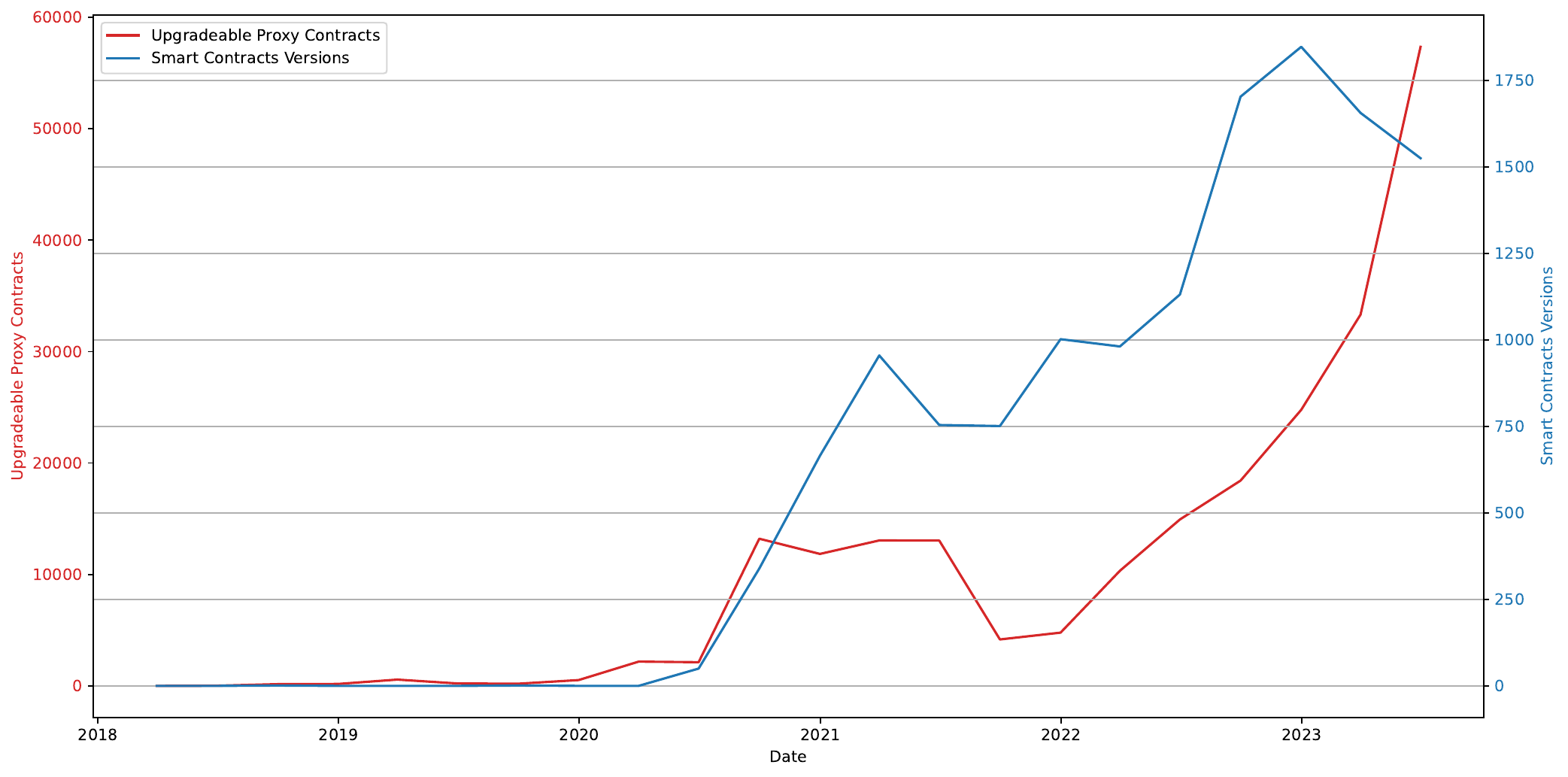}
    \caption{Comparative trends of upgradeable proxy contracts and smart contract versions}
    \label{fig:RQ2D}
\end{figure}
\begin{observation}\textbf{The trend in smart contract versions shows an increasing trajectory, closely related to the trend in upgradeable proxy contracts.}

Figure \ref{fig:RQ2D} presents a comparison trend analysis between the deployment of upgradeable proxy contracts and the progression of smart contract versions over time, utilizing a dual-axis approach due to the significant numerical difference between the two datasets. The x-axis represents the timeline by quarters, illustrating the progression over several years. The primary y-axis on the left, colored in red, represents the count of upgradeable proxy contracts. 
A secondary y-axis on the right, highlighted in blue, details the count of smart contract versions. As in Figure \ref{fig:RQ1B}, the last quarter's data was scaled due to incomplete data.

Initially, the figure illustrates a period where the number of upgradeable proxy contracts is very low, correlating with an absence of smart contract versions. This early stage reflects the foundational need for upgradeable proxies to facilitate contract upgrades, making the initial scarcity of versions expected, given the minimal deployment of upgradeable proxies.
As the timeline progresses, both trends begin to exhibit a parallel increase. The rise in upgradeable proxy contracts, after a certain lag period, is followed by an increase in the number of smart contract versions. This sequence underscores the process wherein enhancements and modifications to smart contracts, captured as versions, are inherently linked to the availability and use of upgradeable proxies. The observed lag between these increases is a natural aspect of the development cycle, accounting for the time required to implement, test, and deploy contract upgrades.

A specific observation in the period after mid-2021 shows a temporary decline in the deployment of upgradeable proxy contracts. Interestingly, this does not lead to an immediate decrease in smart contract versions. Instead, an increase in versions is recorded, likely in response to the preceding rise in upgradeable proxies. This demonstrates the delayed effect of proxy deployment on contract versioning, highlighting the time-dependent nature of contract upgrades in response to earlier increases in upgradeable proxies. 

The final drop observed in the number of smart contract versions towards the end of the period analyzed can be attributed to a reduction in the number of deployed contracts. This observation might initially suggest a decrease in activity or interest. However, this decrease could reflect a recurring pattern seen in previous periods, where specific months showed a similar drop in deployment data. Such patterns indicate that the observed decline might not signify a long-term decrease in smart contract deployments. Instead, it could represent cyclical or seasonal variations in contract creation activities, suggesting a normal fluctuation rather than a diminishing trend in the development and deployment of smart contracts.
\end{observation}

\begin{mdframed}[linewidth=1pt, linecolor=black]
% \RaggedRight
\textbf{RQ2. How likely is an upgradeable contract to be upgraded?} 

\vspace{.5ex}
\noindent\textbf{Answer:} The results show a cautious yet evolving approach towards smart contract upgrades within the community. Despite the inherent technical capabilities for frequent updates, actual upgrade practices are less common, likely due to the complexities and risks associated with post-deployment modifications. The strategies range from dynamic, frequent updates to more conservative, stability-focused versioning, reflecting different contracts' diverse needs and contexts. In essence, the decision to upgrade depends on a balance between technical feasibility, risk management, and specific operational requirements. A temporal analysis shows an increasing trend in both the deployment of upgradeable proxy contracts and the progression of smart contract versions over time. This trend illustrates a natural development cycle where the availability and use of upgradeable proxies facilitate the iterative enhancement of smart contracts, evidenced by the parallel growth in the number of versions.
\end{mdframed}

\subsection{RQ3: Analyzing Post-Upgrade Changes in Smart Contracts}
\afterpage{
\begin{figure}
    \centering
    \includegraphics[width=\textwidth]
    {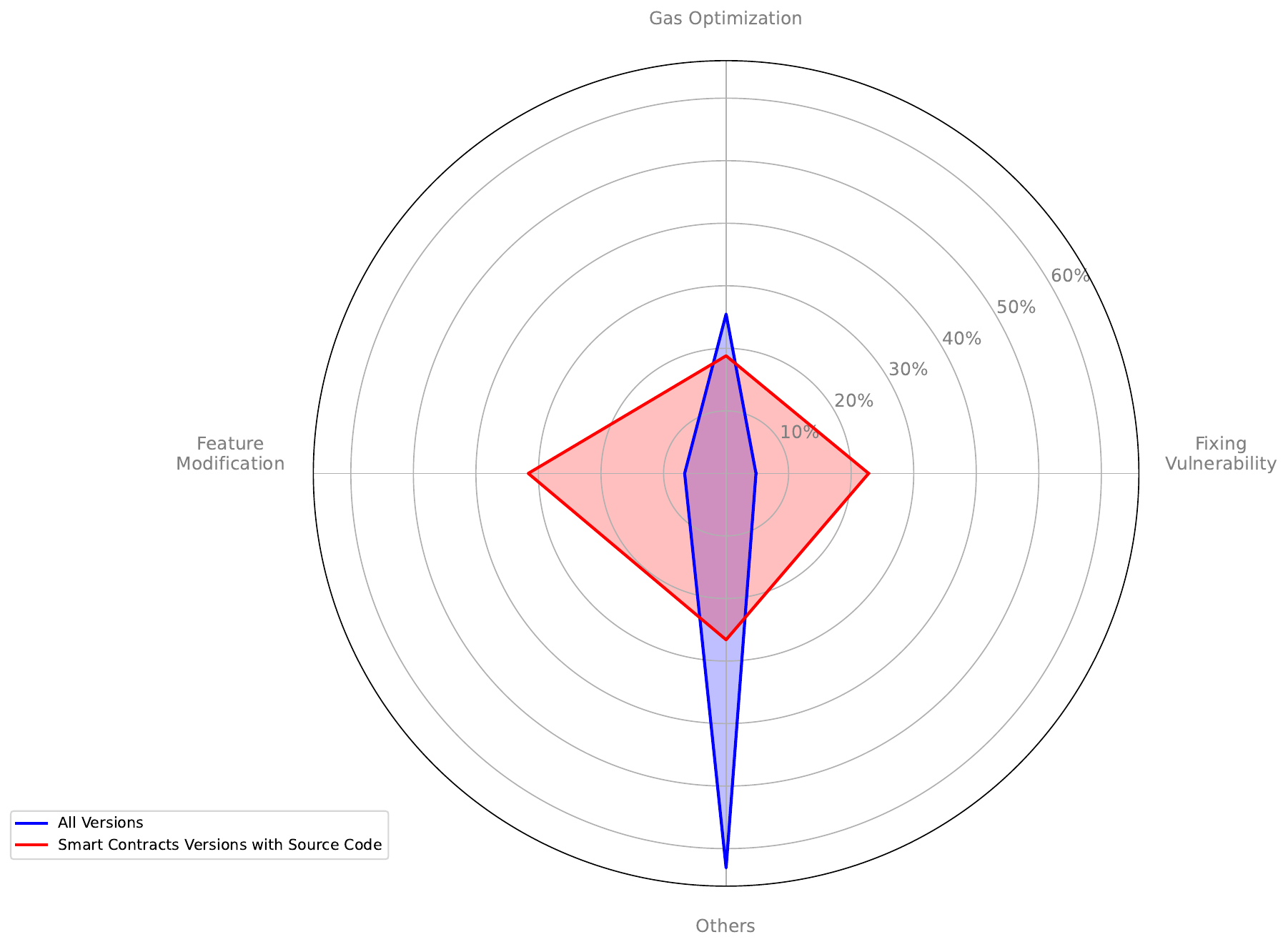}
    \caption{Post-Upgrade changes in smart contracts (RQ3)}
    \label{fig:RQ3}
\end{figure}}
RQ3 aims to identify post-upgrade changes in smart contracts and categorize them as fixing vulnerabilities, gas optimization, feature modification, or others.

Figure \ref{fig:RQ3} is a radar chart representing upgrade patterns' distribution in two datasets: All Versions and Smart Contracts Versions with Source Code. It has four axes, each corresponding to a different post-upgrade change category. The axes start from a common central point and extend outward, equally spaced. The values are represented as percentages of the total counts in each category.
Each dataset is plotted as a closed loop, forming a shape visually representing each pos-upgrade change's relative emphasis. The All Versions dataset is represented in blue, which includes all smart contract versions identified in RQ2. The Smart Contracts Versions with Source Code dataset is in red, focusing on only smart contract versions with available source codes. 

\refstepcounter{observation}
\begin{observation}\textbf{A notable variety in the post-upgrade modifications is observed, with a significant portion categorized as \textit{Others}, suggesting a range of modifications beyond the primary identified categories and the impact of unavailable source code in some instances.} Reflecting on the radar chart, we see a clear distinction in the upgrade patterns between all smart contract versions and those with accessible source code. In the dataset representing all versions, the others category is predominant, consisting of 3,474 instances, accounting for 62.5\% of the upgrades. This suggests that many modifications extend beyond the primary categories identified, and also, in some cases, the lack of source codes for these upgrades contributes to this high percentage. Gas optimization follows with 1,403 instances, approximately 25.3\% of the dataset.
\end{observation}

\begin{observation}\textbf{The availability of source code shifts the focus toward feature modification and Fixing Vulnerability, highlighting the role of accessible code in understanding upgrade causes.} The dataset for smart contracts with source code demonstrates a balanced distribution of post-upgrade changes'. Feature modification has 365 instances, representing 31.68\% of the dataset, indicating a strong focus on enhancing and evolving contract features. On the other hand, fixing vulnerability presents 22.83\% of the modifications, illustrating a significant commitment to maintaining contract security. Both of these categories rely heavily on the availability of the source code. Hence, it has the same percentage in All versions of the dataset. The feature modification category consists of introducing new features or deleting existing features, where each has the following percentages: 95.07\% and 29.59\%, respectively. Within a single upgrade, it is possible to have both new features introduced and existing features deleted. This overlap explains why the combined percentages for new and deleted feature modifications exceed 100\%. 
Gas optimization is present in about 18.84\% of the dataset, a substantial but relatively smaller proportion compared to all versions. The others category sees a significant reduction to 307 instances, representing 26.65\% of this dataset, indicating that the lack of source code contributed to the prevalence of this category.
\end{observation}

\begin{observation}\textbf{Upgrades addressing multiple aspects (security, features, and gas optimization) simultaneously are rare, underlining such upgrades' complexity and resource intensity.} We noticed in both datasets the existence of the combinations of these categories, where a single upgrade showed the existence of more than one changes. The pairing of fixing vulnerability with feature modification, particularly introducing new features, stands out, addressing approximately 18.2\% of the dataset, where developers are concurrently enhancing security and adding new functionalities. In the dataset for smart contracts with source code, the co-occurrence of the three specific post-upgrade changes (fixing vulnerability, feature modification, and gas optimization) is rare. These combined modifications are present in 59 instances, constituting approximately 5.12\% of the total upgrades. This low percentage highlights the infrequency of such comprehensive upgrades within a single iteration. Concurrently integrating all these modifications (enhancing security, introducing or modifying features, and optimizing for gas efficiency) in one update cycle is a complex and less common practice. This rarity underlines the challenges and potential resource intensiveness of implementing multifaceted upgrades in smart contracts.
\end{observation}

\begin{mdframed}[linewidth=1pt, linecolor=black]
\textbf{RQ3. What Changes Occur in Smart Contracts Post-Upgrade?} 

\vspace{.5ex}
\noindent\textbf{Answer:} The post-upgrade changes in smart contracts versions, as analyzed through both comprehensive and source code-specific datasets, predominantly focus on diverse motivations (others) in the absence of source code and shift towards a more balanced emphasis on feature modification and security enhancement when source code is available. The rarity of comprehensive upgrades involving simultaneous security, feature, and efficiency improvements highlights the complex and resource-intensive nature of such endeavors in smart contract development.
\end{mdframed}
\subsection{RQ4: Upgradeability Impact on Contract's Activity level}

As discussed in Section ~\ref{SRQ4}, we employed a regression model to evaluate the relationship between contract upgrades (versions) and usage (transaction numbers). In this model, transaction numbers were set as the dependent variable, while the version lifespan was the independent variable. This approach aimed to account for variances in activity levels attributable to the different lifespans of smart contract versions.
\refstepcounter{observation}

\begin{observation}\textbf{A significant portion of contracts (90.16\%) recorded only one transaction, usually associated with their creation, pointing towards a prevalence of non-active contracts.} Interestingly, most contract versions recorded only one transaction (90.16\%), typically associated with their creation. This pattern suggested a predominantly non-linear relationship between the number of transactions and the contract versions, casting doubt on the suitability of a linear regression model for this data. 
Since analyzing the dataset through regression models provides valuable insights into the relationship between contract upgrades, represented by 'Lifespan' and 'Version,' and their usage in terms of 'Total Transactions.'
\end{observation}

\begin{observation}\textbf{The data suggests a non-linear relationship between contract versions and transaction numbers.} Initially, we applied a linear regression model to the data. The coefficients obtained were approximately \(5.65 \times 10^{-7}\) for 'Lifespan' and \(0.0611\) for 'Version'. These coefficients indicate that an increase in either lifespan or version is associated with an increase in total transactions, with the specified magnitudes. The intercept of the model was around \(-1.16\), marking the point where the regression line would intersect the y-axis.

However, the model's fit to the data, as indicated by the Mean Squared Error (MSE) and the R² Score, was less than ideal. The MSE was quite high at \(3813.33\), and the R² Score was \(-0.03\), suggesting that the linear model did not adequately capture the relationship between the variables. The negative R² score is particularly telling, as it implies that the model may not be the best fit for this dataset or it fails to account for other influential factors.

Figure \ref{fig:RQ4-A} displays the relationship between actual and predicted transaction numbers using a Linear Regression model. In this plot, the x-axis represents the actual transaction numbers, while the y-axis denotes the predicted transaction numbers by the model. A dashed line across the plot symbolizes perfect prediction accuracy, where the predicted values exactly match the actual values. The scatter points illustrate the correlation between the Linear Regression model's predictions and the actual transaction numbers. The proximity or dispersion of these points from the dashed line indicates the model's accuracy, with a closer alignment suggesting higher accuracy and a broader spread indicating variance.

A notable observation from Figure \ref{fig:RQ4-A} is the wide spread of scatter points around the dashed line, indicating a significant variance between the Linear Regression model's predictions and the actual transaction numbers. This spread reflects the model's performance, as quantified by its Mean Squared Error (MSE) and R² Score, underscoring that the predictive accuracy of the Linear Regression model is not optimal. Moreover, a concentration of points around the y = 0 line suggests a trend where the model's predictions tend to be consistently close to zero. This phenomenon indicates potential issues with the model's performance. When numerous data points align closely with the y = 0 line, it may signify either an oversimplified model incapable of capturing the intricacies of the data, or the presence of features that inadequately represent the underlying patterns. Moreover, such clustering around y = 0 could highlight a data imbalance, where a significant portion of the target variable has values close to zero. 

Hence, to thoroughly assess these possibilities, we address the first possibility by implementing a more complex model, specifically a Random Forest, to evaluate whether there is any improvement in predictive performance. This approach allows us to explore whether the increased complexity of the model can better capture the underlying patterns in the data and lead to more accurate predictions.

\begin{figure}
    \centering
    \includegraphics[width=\textwidth]{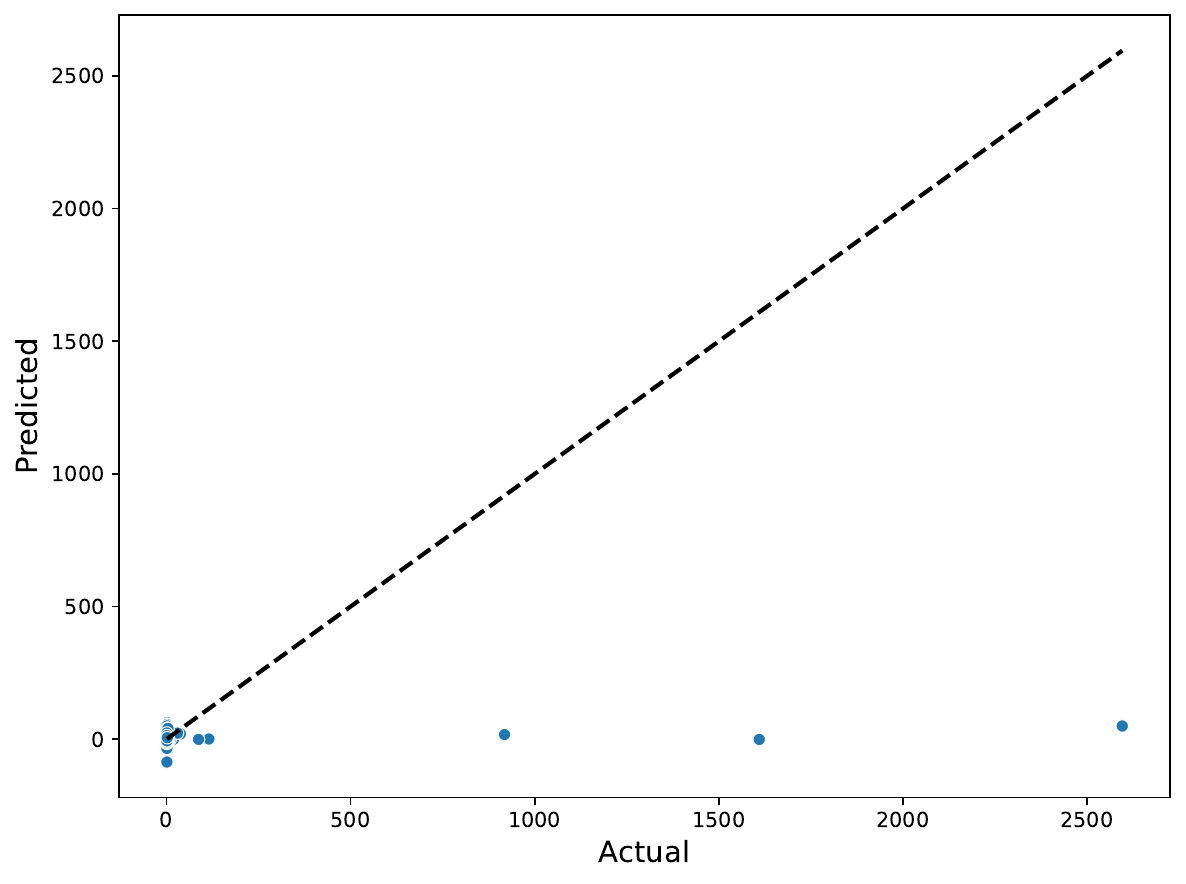}
    \caption{Linear Regression: actual versus predicted}
    \label{fig:RQ4-A}
\end{figure}
\end{observation}

\begin{observation}\textbf{Both linear and Random Forest (non-linear) models show limitations in capturing the relationship between upgrades and activity, indicating that factors beyond the age and version number influence contract activity.} We employed the Random Forest Regressor, a non-linear model for effectively handling complex datasets with multiple variables in response to these limitations. The Random Forest model showed a performance improvement. The MSE reduced to approximately \(2378.09\), indicating a better fit of the model to the data. Additionally, the R² Score increased to about \(0.358\), suggesting that this model could explain around 35.8\% of the variance in the transaction numbers. This improvement is a positive indicator, yet the R² Score still points to a significant portion of variance unexplained by the model.

Figure~\ref{fig:RQ4-BB} showcases the outcomes using a Random Forest model, a non-linear approach to prediction. The axes are defined similarly to Figure~\ref{fig:RQ4-A}, with the actual transaction numbers on the x-axis and the predicted numbers on the y-axis. The presence of the dashed line continues to represent the goal of perfect prediction accuracy. The scatter points in Figure~\ref{fig:RQ4-BB} are similar to those in Figure~\ref{fig:RQ4-A}. They illustrate the correlation between the model's predictions and the actual transaction numbers. 

In Figure~\ref{fig:RQ4-BB}, the scatter points exhibit a wide distribution around the dashed line, indicating a significant variance between the model's predictions and the actual transaction numbers. Notably, there are more points clustered around or in close proximity to the dashed line compared to Figure~\ref{fig:RQ4-A}. While this may suggest relatively better performance for the regression model in terms of capturing certain trends, the substantial spread of points highlights the model's inability to accurately predict transaction numbers across the entire dataset.  These results, especially the improved performance of the Random Forest model, imply a more complex relationship between contract versions, their age, and the total number of transactions than what a simple linear model can capture.
However, the concentration of points around the y = 0 line is similar to the previous model, which also prompts consideration of other potential factors such as data imbalance. Since a large proportion of contracts in the dataset have only one transactions, this concentration may be influenced by such data imbalance and the presence of other confounding factors.

To explore these possibilities further, we focused on the top 20 proxy contracts with the highest total transactions, as shown in Table~\ref{tab:RQ4}. This approach aims to examine whether distinct patterns emerge among the most active contracts that could be better captured by our models.

\begin{figure}[ht]
\centering
\includegraphics[width=\textwidth]{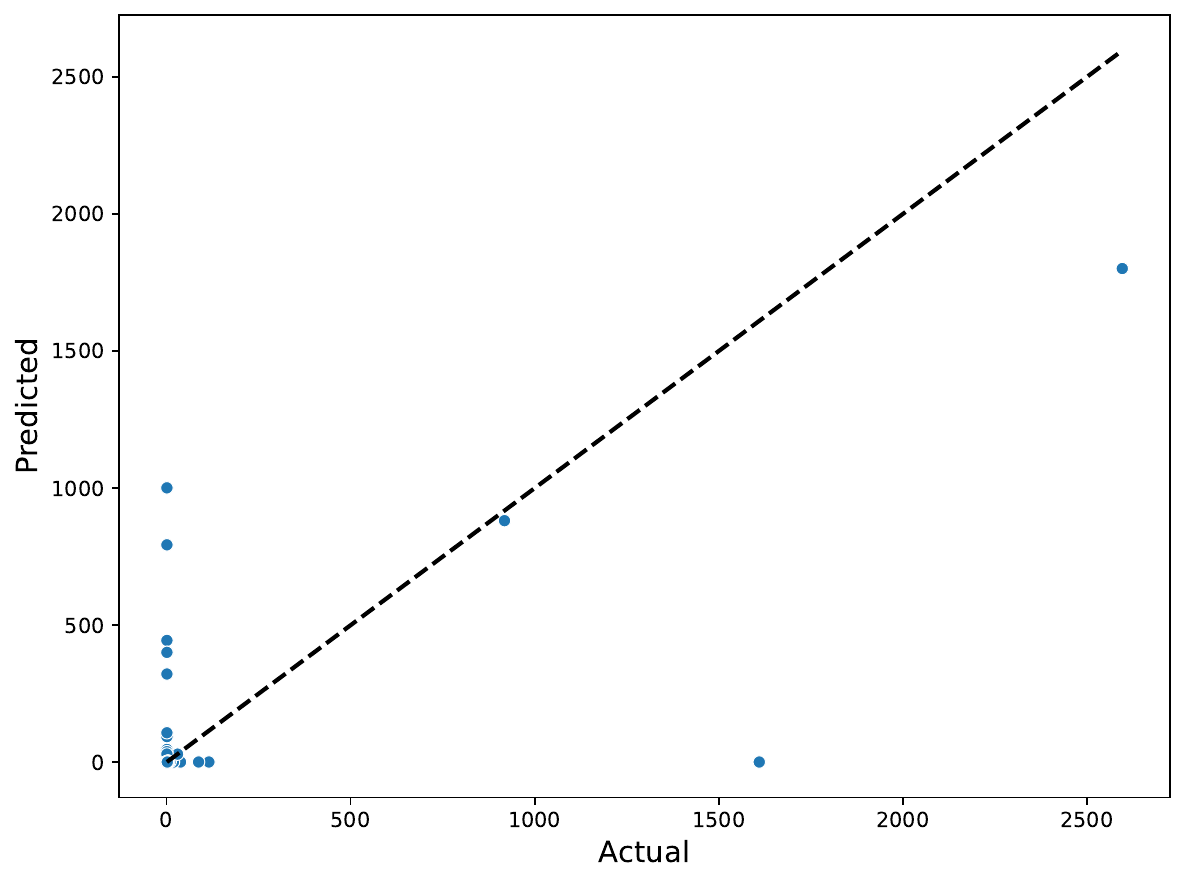}
    \caption{Random Forest (non-linear): actual versus predicted}
    \label{fig:RQ4-BB}
\end{figure}
\end{observation}

\begin{table}[t]
\centering
\caption{Top 20 proxies by total transactions}
\begin{tabular}{lcc}
\hline
\textbf{Proxy Address}                                & \textbf{Number of Versions} & \textbf{Total Transactions} \\ \hline
0x053d938a4d2a6df86d837d66a037444d7bacf3b9 & 2                  & 10,001             \\
0x14f167abdba026c379142436a68d8979a342ecb5 & 3                  & 765                \\
0x1950bb2f0732a78b98adab118791b997a39bb29b & 6                  & 20,070             \\
0x1d3165b32897006935c482ffa0402ad68995ae4c & 5                  & 10,004             \\
0x25f2f80d9a45b641bef25342a1b2a0ae48f78539 & 8                  & 7,218              \\
0x2a53c3fd708d671de8ac07a97456971f2244bd96 & 2                  & 918                \\
0x2a5e45543ecf94de0cc1b574a41b73ff6d456e45 & 1                  & 10,000             \\
0x3570fed1bcfdda4e95cc7038d4d839c06da7e20d & 2                  & 918                \\
0x36e4ba1baec99e7f5950383beda589d3073c4d24 & 2                  & 328                \\
0x3b73c1b2ea59835cbfcadade5462b6ab630d9890 & 6                  & 401                \\
0x3cd5334eb64ebbd4003b72022cc25465f1bfcee6 & 2                  & 2,597              \\
0x4dff845c40d31b3a9b164ae1877901726fa06a2e & 2                  & 918                \\
0x50fda034c0ce7a8f7efdaebda7aa7ca21cc1267e & 2                  & 663                \\
0x537907e5a708c6e2ac607df468ee49111f357596 & 19                 & 10,016             \\
0x59863022b862db70d3c52a1e6d6c0f778763a605 & 6                  & 930                \\
0x5d30ad9c6374bf925d0a75454fa327aacf778492 & 16                 & 238                \\
0x62faa8937f71b4896f9b250f675ff89a5f6875cc & 50                 & 280                \\
0x644b05a51630cd0152ce3bd3fe58bc7763756a2e & 2                  & 918                \\
0x71598610b7713d0321f70662a85a0f95df57db12 & 4                  & 1,836              \\
0xfc3bd18947b719e61b29d58d83bc55e99c7f5b31 & 2                  & 1,612              \\ \hline
\end{tabular}
\label{tab:RQ4}
\end{table}

\begin{figure}[ht!]
    \centering
    \includegraphics[width=\textwidth]{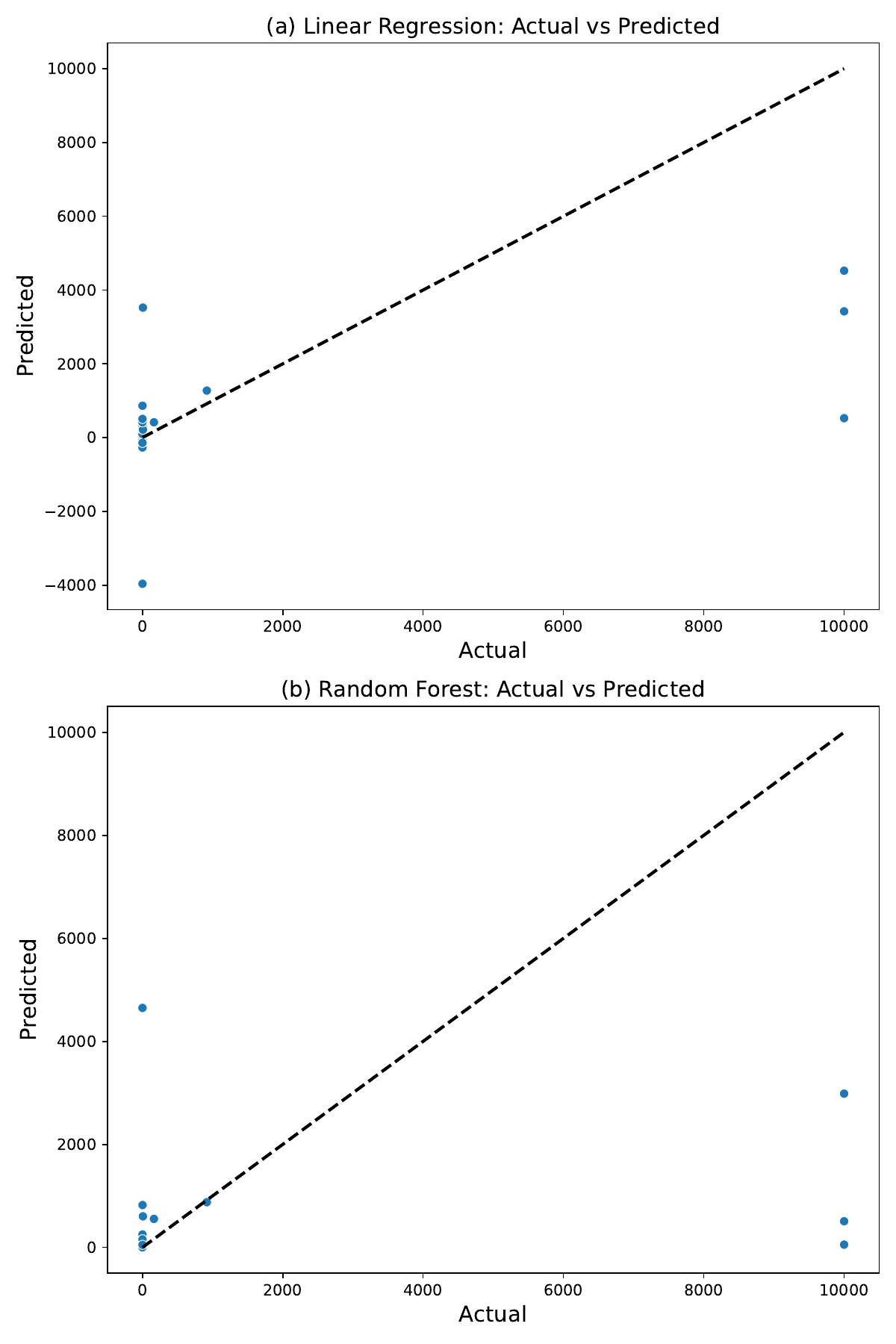}
    \caption{Comparative analysis of regression models on smart contract data for top 20 proxy}
    \label{fig:RQ4-B}
\end{figure}

In analyzing the top 20 proxy contracts, linear and Random Forest regression models were again employed, as shown in Figure \ref{fig:RQ4-B}. 
Figure~\ref{fig:RQ4-B} consists of two subfigures illustrating the performance of different models in predicting transaction numbers for the top 20 proxy contracts. Subfigure~\ref{fig:RQ4-B}(a) displays the relationship between actual and predicted transaction numbers using a linear regression model, while Subfigure~\ref{fig:RQ4-B}(b) shows the relationship for the random forest model. Similar to Figure~\ref{fig:RQ4-A}, the x-axis represents the actual transaction numbers, while the y-axis denotes the predicted transaction numbers by the model. 

The performance of these models was only moderately successful in explaining the variance in transaction numbers. The Random Forest model, in particular, showed a slight performance improvement, indicating some level of non-linear relationship but still pointing towards unaccounted factors influencing contract activity. However, the fact that a substantial portion of variance remains unexplained suggests that there are other factors influencing transaction numbers that are not included in the model. These could be factors intrinsic to the nature of smart contracts, user behaviors, market conditions, or other external variables not accounted for in the current model. 
\begin{mdframed}[linewidth=1pt, linecolor=black]
\textbf{RQ4. How does smart contract upgrading impact the activity level of the contract?} 

\vspace{.5ex}
\noindent\textbf{Answer:} The results suggests that the relationship between contract versions, their lifespans, and transaction numbers is complex and potentially influenced by factors beyond just the age and version number. These could include the contract's purpose, the nature of upgrades, user adoption rates, and external market conditions. The results also showed that the majority of the contracts are non active contract with only one transaction which is the creation transaction of the contract.
\end{mdframed}

\section{Discussion}
\label{Dis}
This section further discusses the RQs results and gives insights into smart contract upgradeability.  
\subsection{RQ1: Prevalence of  Upgradeable Smart Contracts}
Based on the analysis of the prevalence of upgradeable smart contracts in RQ1, We analyzed the community discussions about these standards to reason about the adoption of one standard compared to others. 
We observed that only three standards, Diamond (EIP-2535), Proxy Storage Slots (EIP-1967), and OpenZeppelin, remain actively used. The transparent standard (EIP-1538) was withdrawn with subsequent modification, and the Diamond standard was introduced, indicating a refinement in upgradeable contract approaches. Meanwhile, UUPS (EIP-1822) and EIP-897 standards are stagnant, having not seen significant activity in the past six months. 
From the official discussion of the UUPS (EIP-1822) standard,\footnote{\url{https://ethereum-magicians.org/t/eip-1822-universal-upgradeable-proxy-standard-uups/2842/30}} we also identified a security concern that revolves around the proxiableUUID mechanism. This mechanism is designed to ensure compatibility between a proxy and its implementation contracts by utilizing a unique identifier. The core of the security issue lies in the potential for proxies to mimic the logic intended for implementation contracts, thereby creating a vulnerability during the upgrade process. This vulnerability poses significant risks, such as creating loops of recursive delegation that never reach actual implementation logic, leading to gas depletion and failed transactions. Moreover, it might result in the execution of unintended or malicious logic, compromising contract integrity and potentially leading to the irreversible loss of contract functionality and assets.
This security concern is likely a contributing factor to the observed decline in UUPS (EIP-1822) usage shown in Figure \ref{fig:RQ1B}. The discussion of this issue was in March 2021, and its alignment with the pattern's decline in the figure strongly suggests a correlation between the security concerns and the reduced adoption of this standard in mid-2021. 

From the results of RQ1 in Figure \ref{fig:RQ1A}, we focused on a subset of smart contracts whose proxy types are \textit{Other upgradeable proxy contract}, aiming to classify and understand their underlying structures. We identified three uniform patterns among a subset of these contracts through a detailed analysis:
\begin{itemize}
    \item CErc20Delegator: Identified in 571 contracts, this pattern is part of the Compound protocol and is unique for its detailed constructor, which initializes DeFi-specific parameters (such as the underlying asset, interest rate models, and initial exchange rates) and sets the implementation contract. The contract emits a \textit{NewImplementation} event when a contract is upgraded, including the old and new addresses of the contract. 
    \item Proxyable/ProxyERC20: Designed for ERC20 token upgradeability, this pattern extends the \textit{Owned} contract (for ownership and access control management) and is marked as `payable.` The constructor of this contract type ensures that the contract cannot be instantiated directly, enhancing its security. The \textit{TargetUpdated} event is used to signal updates to the proxy's target, providing transparency for contract upgrades. This pattern was found in 19 contracts from the \textit{Other upgradeable proxy contract} category.
    \item Custom Proxy Contracts on Polygon: This proxy, appearing in 31 contracts, features a \textit{ProxyUpdated} event and suggests the deployment of custom proxy contracts on the Polygon network. Although they align with ERC-897's DelegateProxy pattern, they are recognized under a distinct interface, \textit{IERCproxy}.
\end{itemize}
\subsection{RQ2: Historical Versions of Smart Contracts}
The findings from RQ2 reveal that only 0.34\% of upgradeable smart contracts are actually upgraded after deployment. This observation challenges the presumed advantage of upgradeability, highlighting a disparity between theoretical capabilities and their application in practice. Two primary factors could explain this phenomenon: the contracts might be of high initial quality due to thorough auditing and testing before deployment, or the complexity and risks associated with upgrading might deter their application.

Firstly, the high initial quality of these smart contracts can be attributed to rigorous auditing and testing procedures prior to deployment. Auditing involves a detailed examination of the contract's code by security experts to identify potential vulnerabilities, while testing, encompassing both automated and manual methods, ensures the contract operates as intended under various scenarios. This scrutiny level helps minimize the need for future upgrades by ensuring the contract is as secure and functional as possible from the outset. This perspective aligns with the industry's growing emphasis on comprehensive audit practices for smart contracts, recognizing the critical importance of security in decentralized finance (DeFi) and other blockchain applications.

On the other hand, the complexity and inherent risks of the upgrade process itself could prevent developers and contract owners from initiating upgrades. Upgrading a smart contract involves technical challenges, such as ensuring compatibility with existing functionalities and managing the contract's state during the transition, as well as operational risks. Moreover, every new code deployment risks introducing unforeseen vulnerabilities, making the upgrade process a cautious endeavor that requires meticulous planning, testing, and auditing similar to the initial deployment.

Furthermore, the impact of upgrades on the quality of smart contracts is nuanced. While the ability to upgrade allows for the correction of flaws, adaptation to new requirements, and introducing improvements, it also introduces a layer of uncertainty. Each upgrade carries the potential to enhance the contract's functionality and security but also risks compromising its integrity if not carefully executed. Therefore, the decision to upgrade involves weighing the benefits of improved functionality and security against the risks of introducing new vulnerabilities and the operational complexities involved in the upgrade process.

\subsection{RQ3: Analyzing Post-Upgrade Changes in Smart Contracts}
Based on the results from RQ3,  we investigated the relationship between the availability of source code and the upgrade frequency of smart contracts. Our analysis divides smart contracts into two groups based on source code availability, leading to a comparative study of how this factor influences the evolutionary trajectory of these contracts.

Contracts with available source code exhibit a higher mean number of versions than those without. Specifically, the mean version count for contracts with accessible source code is approximately 2.71. In contrast, contracts with no available source code have a lower mean version count, around 2.00. This statistical disparity highlights the difference in upgrade frequency and suggests a broader trend toward more dynamic development cycles in open-source contracts. This pattern reflects the broader benefits of open-source practices, including enhanced community engagement, increased transparency, and a more iterative approach to development. These factors collectively contribute to a more responsive development ecosystem for smart contracts, underlining the importance of source code availability in fostering the growth and security of blockchain technologies.

\subsection{RQ4: Upgradeability Impact on Contract's Activity level}

Our study aimed to investigate the relationship between smart contracts version activity level, measured by total transactions, and their upgrade frequency. Since there is no linear relationship between the number of transactions and the age of the contract, we focused on comparing the upgrade intervals of the most actively used contracts with those less active. This approach was intended to identify if activity level influences upgrade practices, as we have noticed many smart contract updates within minutes.
We divided the proxy contracts into two groups for analysis: the top 20 proxies by transaction volume and the remaining proxies, excluding the latest version of each, to consider only complete upgrade cycles. The top 20 proxies demonstrated an average upgrade interval of approximately 609 days (20.3 months), while the remaining proxies had a shorter interval of about 521 days (17.4 months).

This differential suggests that more actively used contracts are upgraded less frequently. This could imply that high transaction volumes lead to a preference for stability and reduced upgrade frequency, possibly due to the complexities of updating heavily used systems. Alternatively, it might reflect a strategic decision to minimize disruptions in high-stake environments.

\subsection{Smart Contract Upgradeability}
The empirical findings shed light on the ``to upgrade or not to upgrade'' dilemma faced by the smart contract community. This dilemma is rooted in the contrast between blockchain's foundational principle of immutability and the practical need for adaptability to address evolving requirements and unforeseen vulnerabilities. The finding that only 3\% of smart contracts are designed as upgradeable proxy contracts suggests a cautious approach within the community toward embracing upgradeability. This caution may arise from concerns over the complexities and potential risks associated with managing upgradeable contracts, such as introducing new vulnerabilities and the operational challenges of executing upgrades, though the exact reasons are not definitively known.

Furthermore, the observation that a small percentage of these upgradeable contracts have undergone upgrades at least twice underscores a broader hesitancy to utilize the upgrade capability, even when technically feasible. This hesitancy could indicate a preference for stability and predictability over the potential benefits of iterative improvements. The low frequency of upgrades among upgradeable contracts hints at apprehensions about the impacts of such changes, potentially including technical risks and implications for user trust and contract reliability.

This scenario reveals a complex landscape in blockchain development, where decisions around making a smart contract upgradeable and subsequently acting on this capability involve weighing multiple considerations. Developers and stakeholders seem to navigate a delicate balance between addressing critical vulnerabilities, adapting to new requirements, and maintaining the trust and integrity of smart contract immutability. The balance between pursuing innovation and ensuring stability is central to the upgrade dilemma, reflecting wider challenges in the evolution of blockchain technologies and their applications.

\section{Related Work}
\label{RW}
There are only a handful of studies focused on the upgradeability of smart contracts. This section presents and discusses these existing studies for smart contract upgradeability. 

Salehi et al.~\cite{salehi2022not} analyzed and evaluated smart contract upgradability patterns. The authors presented a framework for measuring the number of upgradeable Ethereum contracts which utilize certain well-known upgradeable proxy patterns. Furthermore, they have analyzed how access control is implemented over smart contract upgradeability.

Bui et al.~\cite{bui2021evaluating} proposed a Comprehensive-Data-Proxy pattern to upgrade smart contracts while enhancing security resilience and scalability. The authors investigated three popular Ethereum attacks that affect smart contract upgradeability: cross-function Reentrancy, typical Reentrancy, and DAO attacks. The presented pattern improves resilience against such attacks compared to the previous upgrading approaches.

Chen et al.~\cite{chen2020finding} introduced a deep learning-based method to detect security issues in the updated version of a destructed smart contract. A contract can be destroyed on Ethereum only by using the Selfdestruct function, which transfers all the Ethers on the contract balance to the contract owner. The authors compare the historical version of destructed contracts and investigated whether security issues were detected in the destructed contracts. 

Fröwis et al.~\cite{frowis2022not} investigated the impact and evaluated the CREATE2 instruction adoption in Ethereum smart contracts. The CREATE2 instruction allows the contract to be modified after deployment on a given address. Furthermore, the authors identified several use cases and attack vectors for the CREATE2 instruction.

Bodell III et al.~\cite{bodell2023proxy} conducted research on upgradeable smart contracts using proxies to investigate the current situation and identify security issues. The authors created a comprehensive classification of proxy-based USCs that can distinguish their behaviors based on both structural and semantic characteristics. They also developed USCHUNT, a static analysis framework for detecting and examining USCs, and used it to analyze over 800K smart contracts across eight popular blockchains. Their findings include 11 distinct USC design patterns and six types of security and safety concerns.

Ebrahimi et al.~\cite{ebrahimilarge} investigated the proxy pattern (whether upgradeable or non-upgradeable) using a dataset containing 50 million smart contracts. Using a behavioral detection technique, they discovered that over 14\% of the contracts are active proxies. The study also revealed an increasing trend in using proxy contracts throughout Ethereum's history, highlighting their significant role in enhancing modularity. Moreover, the authors identified 12 distinct creational patterns for deploying proxies (upgradeable and non-upgradeable), which we categorized as off-chain and on-chain styles based on where the deployment script operates. However, this classification does not specifically focus on upgradeable proxy contracts and deviates from established standard classifications.

Table \ref{tab:RW} illustrates the comparison between previous studies and our empirical study regarding smart contract upgradeability. The focus of existing studies on smart contract upgradeability has primarily been detecting upgrading approaches, particularly proxy-based methods. However, only two studies~\cite{salehi2022not,bodell2023proxy} provided insights on the prevalence of upgrading proxy patterns. In this study, we focused beyond the prevalence of upgrading contracts (RQ1), where we analyzed smart contract versions (RQ2), post-upgrade modifications (RQ3), and the impact of upgrading on the adoption of smart contracts (RQ4). Nevertheless, there are aspects that this study did not cover which were discussed in previous studies, such as the security issues related to the existing proxy approaches~\cite{bodell2023proxy} as well as the access control analysis of these contracts~\cite{salehi2022not,ebrahimilarge}. While these are interesting topics, they were not within the scope of our empirical study on smart contract upgradeability.

\begin{table}[t]
\caption{Comparison of studies on smart contract upgradeability}
\label{tab:RW}
\resizebox{\textwidth}{!}{%
\begin{tabular}{@{}ccccccc@{}}
\toprule
Study &
  \begin{tabular}[c]{@{}c@{}}Smart Contract \\ Upgrading Approach\end{tabular} &
  \begin{tabular}[c]{@{}c@{}}Smart Contract \\ Versions\end{tabular} &
  \begin{tabular}[c]{@{}c@{}}Post-Upgrade \\ Analysis\end{tabular} &
  \begin{tabular}[c]{@{}c@{}}Security \\ Issues\end{tabular} &
  \begin{tabular}[c]{@{}c@{}}Access Control\\ Analysis\end{tabular} &
  \begin{tabular}[c]{@{}c@{}}Activity Level\\ Analysis\end{tabular} \\ \midrule
\cite{salehi2022not}     & Proxy                                                             &            &            &            & \checkmark &            \\ \midrule
\cite{bui2021evaluating} & Proxy                                                             &            &            &            &            &            \\ \midrule
\cite{chen2020finding}   & \begin{tabular}[c]{@{}c@{}}Self-destruct\\  function\end{tabular} &            &            &            &            &            \\ \midrule
\cite{frowis2022not}     & \begin{tabular}[c]{@{}c@{}}CREATE2\\ instruction\end{tabular}     &            &            &            &            &            \\ \midrule
\cite{bodell2023proxy}   & Proxy                                                             &            &            & \checkmark &            &            \\ \midrule
\cite{ebrahimilarge}     & Proxy                                                             &            &            &            & \checkmark &            \\ \midrule
Our Study                & Proxy                                                             & \checkmark & \checkmark & \checkmark &            & \checkmark \\ \bottomrule
\end{tabular}%
}
\end{table}
\section{Threats to Validity}
\label{TH}
This section outlines potential threats to the validity of our study, ranging from issues that might affect the internal validity to those that could impact the external validity.
\subsection{Internal Validity}
A significant concern in our study is the limitations associated with using automated tools for detecting vulnerabilities. These tools could potentially produce false positives or miss vulnerabilities, impacting the accuracy of our findings. We have implemented robust filtering strategies to mitigate this and applied majority rules in our analysis. This approach is designed to minimize the occurrence of false positives. 

Another threat to internal validity stems from identifying and categorizing smart contracts and their versions. While our approach does not involve manual labeling, which could introduce subjective biases, there remains a possibility of inaccuracies in the automated identification and classification process. We have extensively tested and validated each step of our methodology to counter this risk. This testing serves not only as a check on the accuracy of our processes but also as a validation of the methods themselves. It provides us with confidence in the applicability of our findings, ensuring that our conclusions are based on reliable and correctly categorized data.

Moreover, when detecting upgradeable proxy contracts, there is a risk of incorrect identification, which could impact our results. To mitigate this, we have used a well-known EVM proxy detection tool as a base for our detection mechanism. This tool is recognized for its accuracy and reliability in identifying proxies, reducing the risk of misclassification. Moreover, we analyzed the standard description for each proxy type to identify the unique characteristics of each standard and incorporate it within our detection method to ensure more precise identification of upgradeable proxy contracts, thereby minimizing errors in classification.

\subsection{External validity}
Our study focuses on Ethereum smart contracts, which may limit our findings' generalizability to other types of smart contracts or blockchain platforms. While the study aimed not to generalize all smart contracts but to provide insights into the impact of upgrading on contracts' security, we acknowledge this limitation. The thorough testing of sample data ensures a solid foundation for our conclusions within the scope of Ethereum-based contracts.

Additionally, our study specifically focuses on the proxy pattern among upgradeable smart contracts, acknowledging that other patterns exist but are not the focus of our research. We have excluded other types as they do not exhibit unique characterization to distinguish them. Moreover, some types have minimal prevalence, such as data separation, which represents only 0.0007 \% of the dataset. Threats to validity may arise from the limited scope of our study, as we only consider the proxy pattern among upgradeable smart contracts due to its significant role and the concentrated attention it receives within the smart contract development community. Moreover, focusing on proxy patterns allows for a more detailed and impactful analysis of the most commonly employed upgrade techniques in the Ethereum platform. However, this narrow focus may limit the generalizability and applicability of our findings to other patterns or platforms.
\section{Conclusion}
\label{Con}
In this paper, we have provided a comprehensive analysis of upgradeable smart contracts, revealing insights into their prevalence, upgrade patterns, and implications. Our study found that upgradeable proxy contracts represent a small fraction of the total smart contracts, indicating a limited adoption of upgradeable patterns. We also observed that the actual frequency of upgrades in smart contracts is comparatively low despite the availability of technical capabilities for frequent modifications.
Our analysis of the root causes of upgrades showed diverse motivations, with a significant emphasis on feature modifications and security enhancements, especially when source code is available. In examining the impact of upgradeability on the activity level of contracts, we uncovered a complex relationship between contract versions, their lifespan, and usage, suggesting influences beyond mere technical factors.

For future research, a focused investigation into the security impacts of upgrades, particularly in the context of identified vulnerabilities in various upgrade patterns, would be valuable. This could lead to a better understanding of the security landscape of smart contracts and inform more secure development practices.
Furthermore, conducting a survey with smart contract developers and the community could be highly informative. This survey would aim to understand the current issues and challenges developers face in upgrading smart contracts. Insights from such a survey would provide valuable perspectives on the practical difficulties, preferences, and considerations that influence decisions around smart contract upgrades. These findings inform strategies to address these challenges, leading to more efficient and secure approaches to smart contract development.
\bibliographystyle{ieeetr}
\bibliography{SCV}

\end{document}